\documentclass[reprint,aps,prd,amssymb, amsmath, nofootinbib]{revtex4-2}

\usepackage{graphicx}

\usepackage{mathtools}
\usepackage[svgnames]{xcolor} 

\usepackage[colorlinks=True, linkcolor=SlateBlue,
            citecolor=SteelBlue,urlcolor=BlueViolet]{hyperref}
\usepackage[capitalise]{cleveref}
\usepackage{orcidlink}
\usepackage{extpfeil}

\usepackage{aas_macros}

\newcommand{\vect}[1]{\boldsymbol{#1}}

\newcommand{\la}{\langle}
\newcommand{\ra}{\rangle}

\begin{document}

\author{Hang Yu\,\orcidlink{0000-0002-6011-6190}}
\email{hang.yu2@montana.edu}

\affiliation{eXtreme Gravity Institute, Department of Physics, Montana State University,
Bozeman, Montana 59717, USA}

\author{Phil Arras\,\orcidlink{0000-0001-5611-1349}}
\affiliation{Department of Astronomy, University of Virginia, P.O. Box 400325, Charlottesville, VA 22904, USA}

\author{Nevin N. Weinberg\,\orcidlink{0000-0001-9194-2084}}
\affiliation{Department of Physics, University of Texas at Arlington, Arlington, TX 76019, USA}

\title{Dynamical tides during the inspiral of rapidly spinning neutron stars: \\Solutions beyond mode resonance}

\begin{abstract}
We investigate the dynamical tide in a gravitational wave (GW)-driven coalescing binary involving at least one neutron star (NS). 
The deformed NS is assumed to spin rapidly, with its spin axis anti-aligned with the orbit. 
Such an NS may exist if the binary forms dynamically in a dense environment, and it can lead to a particularly strong tide because the NS f-mode can be resonantly excited during the inspiral. 
We present a new analytical solution for the f-mode resonance by decomposing the tide into a resummed equilibrium component varying at the tidal forcing frequency and a dynamical component varying at the f-mode eigenfrequency that is excited only around mode resonance.
This solution simplifies numerical implementations by avoiding the subtraction of two diverging terms as was done in previous analyses. It also extends the solution's validity to frequencies beyond mode resonance. 
When the dynamical tide back reacts on the orbit, we demonstrate that the commonly adopted effective Love number is insufficient because it does not capture the tidal torque on the orbit that dominates the back reaction during mode resonance. An additional dressing factor originating from the imaginary part of the Love number is therefore introduced to model the torque. 
The dissipative interaction between the NS and the orbital mass multipoles is computed including the dynamical tide and shown to be subdominant compared to the conservative energy transfer from the orbit to the NS modes. 
Our study shows that orbital phase shifts caused by the $l=3$ and $l=2$ f-modes can reach 0.5 and 10 radians at their respective resonances if the NS has a spin rate of 850 Hz and direction anti-aligned with the orbit. Because of the large impact of the $l=2$ dynamical tide, a linearized analytical description becomes insufficient, calling for future developments to incorporate higher-order corrections.  
After mode excitation, the orbit cannot remain quasi-circular, and the eccentricity excited by the $l=2$ dynamical tide can approach nearly $e\simeq 0.1$, leading to non-monotonic frequency evolution which breaks the stationary phase approximation commonly adopted by frequency domain phenomenological waveform constructions. 
Lastly, we demonstrate that the GW radiation from the resonantly excited f-mode alone can be detected with a signal-to-noise ratio exceeding unity at a distance of 50 Mpc with the next-generation GW detectors. 
\end{abstract}

\maketitle

\section{Introduction}

The first observation of a binary neutron star (BNS) merger on August 17, 2017,  demonstrated the future promise of detecting imprints of matter effects of a neutron star (NS) in gravitational wave (GW) signals~\cite{GW170817, GW170817prop, GW170817eos, GW170817pg}. 
Detecting an NS with a significant spin misaligned with the orbit can be especially interesting for at least two reasons. First, such a system can be highly informative about the formation history of the binary.
A misaligned spin has been considered to be smoking gun evidence that the binary formed dynamically in a dense stellar environment~\cite{Rodriguez:16}. While most NSs in the Galactic BNS population have low spins~\cite{Zhu:18}, nothing fundamental prevents a rapidly spinning NS in a coalescing binary. The maximum dimensionless spin an NS can have before it breaks up is about 0.67 \cite{Lo:11}, or about a spin rate of 1000\,Hz (depending on the equation of state). 
The fastest-spinning pulsar known, J1748-2446ad, has a spin rate of 716 Hz. It resides in the globular cluster Terzan 5 and has a light companion~\cite{Hessels:06}, both conditions favorable for the dynamical production of GW sources~\cite{Rodriguez:15}. If an NS similar to J1748-2446ad is captured into a compact and/or highly eccentric orbit through, e.g., binary-single encountering~\cite{Samsing:14}, it may merge quickly while retaining a significant spin when entering the sensitivity band of a ground-based GW detector. 
Second, matter effects can also lead to rich dynamics in the binary's evolution when the spin is misaligned \cite{Lyu:23}. It can both modify the precession of the orbit~\cite{LaHaye:22, Divyajyoti:24} and cause extra dephasing of the GW signal through both Newtonian (via tidal interactions~\cite{Lai:93, Lai:94a, Lai:94b, Flanagan:08, Hinderer:10, Bini:14, Bernuzzi:15}) and post-Newtonian (or PN via, e.g., quadrupole-monopole interactions~\cite{Krishnendu:17, Harry:18}) effects. 

Of particular interest to this work is the tidal interaction and we will focus on a special scenario in which the deformed NS has a spin anti-aligned with the orbit. It has been shown in both theoretical studies~\cite{Ho:99, Ma:20, Steinhoff:21, Kuan:22, Kuan:23} and numerical relativity simulations~\cite{Hotokezaka:15, Dietrich:18, Foucart:19} that tidal deformation is significantly amplified when the spin is anti-aligned, because the anti-aligned spin shifts the eigenfrequency of the NS fundamental mode oscillation, or f-mode, down to a lower value in the inertial frame (or equivalently, it shifts the tidal forcing frequency up in the frame corotating with the NS), so that the NS f-mode can be resonantly excited during the inspiral stage. 

While leading to a strong signal, the resonant excitation of a mode also leads to theoretical challenges in modeling this effect. 
Some earlier milestones in this effort include Ref.~\cite{Lai:94c} in which the author computed the asymptotic energy transfer of a resonantly excited mode, and Ref.~\cite{Flanagan:07} where the authors identified the key effect of a resonantly excited mode can be modeled as a jump in the orbit's frequency evolution at the mode resonance. Other works discussing mode resonances in coalescing binaries include Refs. \cite{Ho:99, Ma:20, Kuan:22} for f-modes, Refs. \cite{Lai:94c, Reisenegger:94,  Yu:17a, Yu:17b, Kuan:21, Kuan:21b} for gravity modes, Refs. \cite{Ho:99, Xu:17, Poisson:20, Ma:21, Gupta:21} for Rossby/inertial modes, and Refs. \cite{Tsang:12, Pan:20, Passamonti:21} for interface/crustal modes. 

While a model as simple as a jump in the frequency evolution at the mode resonance may be sufficient in modeling other modes whose tidal coupling is small, for the f-mode that dominates the tidal interaction, more detailed modeling is desired when constructing faithful GW waveform templates for parameter estimation. In particular, such a model should smoothly connect the analysis of f-mode in the adiabatic (zero driving frequency) limit~\cite{Lai:93, Lai:94a, Lai:94b, Flanagan:08, Hinderer:10, Bini:14, Bernuzzi:15} to those that include corrections from finite-frequency effects, or even mode resonance. 
Analytical models for the f-mode's evolution and its orbital back-reaction can be highly valuable for parameter estimation purposes because they can be generated in large quantities with manageable computational costs. The inference of tidal signatures in BNS signals further plays a central role in constraining the equation of state of NS matter~\cite{Read:09, Damour:12, DelPozzo:13, Lackey:15, Andersson:18, Landry:19, Matas:20, Pratten:22}.  An analytical model can also guide the empirical calibration against numerical simulations in the nonlinear regime~\cite{Dietrich:17, Dietrich:19}. Some major steps forward in integrating the f-mode resonance into GW waveform construction under the effective-one body (EOB) framework~\cite{Buonanno:99} include Refs. \cite{Steinhoff:16, Hinderer:16} for non-spinning NSs and Ref.~\cite{Steinhoff:21} for spinning ones. 
The authors proposed to modify the tidal Love number with a frequency-dependent ``dressing factor'', known as the ``effective Love number'' so that the tidal back-reaction on the orbit takes the same form as in the adiabatic limit.
The effective Love number description quickly gained popularity in the literature when modeling NS dynamical tides. See, e.g., Refs. \cite{Andersson:20, Andersson:21, Passamonti:22, Gamba:23, Kuan:23, Mandal:23,Pnigouras:24, Pitre:24}.

However, there are a few aspects to be improved for the analysis in Ref. \cite{Steinhoff:21}. One is that its numerical implementation may not be ideal when mode resonance does occur. The lowest-order finite-frequency correction of the tide has a diverging behavior at mode resonance, which is unphysical because the mode spends only a finite amount of time near resonance~\cite{Lai:94c}. The authors of Ref.~\cite{Steinhoff:21}, following their earlier analyses in Refs. \cite{Steinhoff:16, Hinderer:16}, introduced another diverging term with an opposite sign to cancel the divergence. A similar procedure to remove the divergence is also adopted in Ref.~\cite{Ma:20}. While theoretically accurate, subtracting two diverging terms to get a perfect cancellation can be challenging in numerical implementations. 

Moreover, the evolution of the tidally induced NS mass multipole in the formulation of Ref.~\cite{Steinhoff:21} is accurate only up to resonance, and the effective Love number is adequate in describing the back-reaction only when the tidal forcing frequency is below resonance. In particular, we will show explicitly that the effective Love number fails to describe the torque between the NS and the orbit, which dominates the tidal back-reaction around mode resonance. It is actually what leads to the jump in the waveform's frequency evolution shown in earlier analyses~\cite{Flanagan:07}. The effective Love number is also insufficient in describing the GW radiation due to the coherent interaction of NS and orbital mass multipoles when finite frequency corrections are included. Indeed, a recent analysis in Ref.~\cite{Gamba:23} comparing waveforms generated following Ref.~\cite{Steinhoff:21} found it insufficient in describing systems where f-mode resonance happens, and the effective Love number treatment does not improve the agreement with numerical relativity when compared with models without f-mode resonances~\cite{Bernuzzi:15, Akcay:19}. 

We therefore aim to improve the analysis by Ref. \cite{Steinhoff:21} in describing the NS f-mode resonance including the leading-order correction due to NS spin. 
We will start by enhancing the treatment of the mode amplitude evolution in Sec.~\ref{sec:mode_amp} so that all diverging terms are eliminated analytically for easy numerical implementation. More importantly, we extend the validity of the solution to frequencies beyond mode resonance. 
In Sec.~\ref{sec:love_num}, we examine the relation between NS mass multipoles, the effective Love number, and tidal backreactions on the orbit. We will show that the commonly adopted effective Love number captures only the radial interaction but misses the tidal torque, yet the torque is what dominates the back-reaction around mode resonance. A new dressing factor is therefore introduced to describe the torque. 
It is followed by Sec. \ref{sec:dyn_tide} where we describe the system's evolution throughout the inspiral, including both pre and post-resonance regimes. 
We consider in Sec.~\ref{sec:time_phase_shift} the tidal phase shift in the GW signal using both energy balancing arguments and osculating orbits. The osculating orbits further allow us to examine in Sec.~\ref{sec:ecc} the deviation from the quasi-circular inspiral in the post-resonance regime. We also analyze the detectability of the GW  from the resonantly excited f-mode in Sec.~\ref{sec:GW_f_mode}. Lastly, we conclude and discuss in Sec.~\ref{sec:conclusion}.

Throughout the paper, we adopt geometrical units with $G=c=1$. We will use $M_1$ to denote the mass of the tidally deformed NS and $R_1$ its radius. Together they lead to natural units $E_1\equiv M_1^2/R_1$ and $\omega_1\equiv \sqrt{M_1/R_1^3}$. The mass of the companion is denoted with $M_2$. To describe the orbital dynamics, we further define $M_t=M_1 +M_2$, $\mu = M_1 M_2/M_t$, and $\eta=\mu / M_t$ as the total mass, reduced mass, and symmetric mass ratio. 
As our focus is the tidal interaction, we simplify the analysis by modeling the orbital dynamics at Newtonian order while including the leading order quadrupole GW radiation. A more rigorous analysis should incorporate relativistic effects, as they are crucial in the late inspiral stage.  This would entail replacing our Newtonian Hamiltonian to, e.g., that constructed under the EOB framework, as done in Refs.~\cite{Steinhoff:16, Hinderer:16, Steinhoff:21}. We defer such an upgrade to a future study.

\section{Analytical solution of tidal amplitude}
\label{sec:mode_amp}

We begin our discussion by first examining the evolution of the amplitude of a tidally driven NS mode throughout the inspiral stage. The mode amplitude is directly related to the tidally induced mass multipoles [see later Eq. (\ref{eq:multipole_from_amp})] and is what determines the back-reaction on the orbit, a topic we will discuss in detail in Sec.~\ref{sec:love_num}. 

The motion of a perturbed fluid parcel with Lagrangian displacement $\vect{\xi}$ can be written in the corotating frame to linear order in $\vect{\xi}$ as~\cite{Schenk:02} 
\begin{equation}
    \ddot{\vect{\xi}} + 2\vect{\Omega} \times \dot{\vect{\xi}} + \vect{C} \vect{\xi} = \vect{a}_{\rm ext}, \label{eq:xi_eom}
\end{equation}
where $\vect{\Omega}$ is the spin vector, $\vect{C}$ is a self-adjoint operator describing the hydrodynamic response of the star, and $\vect{a}_{\rm ext}$ is an external acceleration acting on the star. 

We perform a phase-space decomposition of the fluid into eigenmodes as~\cite{Schenk:02}
\begin{equation}
    \begin{bmatrix}
        \vect{\xi} \\
        \dot{\vect{\xi}}
    \end{bmatrix}
    =\sum_a q_a
    \begin{bmatrix}
        \vect{\xi}_a \\
        - i \omega_a {\vect{\xi}}_a
    \end{bmatrix},
\end{equation}
where $q_a=q_a(t)$ is a mode's amplitude whose temporal evolution is to be solved in this section, and $\vect{\xi}_a=\vect{\xi}_a(\vect{r}_{\rm ns})$ is the mode's spatial eigenfunction. We use $\vect{r}_{\rm ns}$ to denote a fluid element's position inside the NS, which should be distinguished from the orbital separation denoted by $r$.
The summation runs over all the radial and angular quantum numbers as well as the signs of the mode's eigenfrequency $\omega_a$ (in the corotating frame). When acting on an eigenmode, the operator $\vect{C}$ satisfies 
\begin{equation}
    \vect{C}\vect{\xi}_a = \omega_a^2 \vect{\xi}_a  + 2i \omega_a \vect{\Omega} \times \vect{\xi}_a.  
\end{equation}
The equation of motion for each mode in the co-rotating frame is given by~\cite{Schenk:02} 
\begin{equation}
    \dot{q}_a + i \omega_a q_a = f_a,
\end{equation}
where
\begin{align}
    &f_a = \frac{i}{\epsilon_a} \la\vect{\xi}_a, \vect{a}_{\rm ext}\ra, \\
    &\epsilon_a = 2\la\vect{\xi}_a , \omega_a\vect{\xi}_a + i\vect{\Omega}\times \vect{\xi}_a\ra. 
\end{align}
For this work, we will focus on the case where the binary moves in the x-y plane in an initially quasi-circular orbit, and $\vect{\Omega}$ is either aligned ($\Omega >0$) or anti-aligned ($\Omega<0$) with the z-axis, the direction of orbital angular momentum. We further consider the case where the fluid is perturbed by the companion's tidal field so $\vect{a}_{\rm ext} = -\nabla U$.

We incorporate the spin as a small perturbation. To linear order, 
the eigenfrequency of a mode $a$ in the corotating frame $\omega_a$ is related to its non-rotating value $\omega_{a0}$ via~\cite{Christensen-Dalsgaard:98, Aerts:10}
\begin{align}
    &\Delta \omega_a \equiv \omega_a - \omega_{a0}  = -
    i\frac{\langle \vect{\xi}_a, \vect{\Omega} \times \vect{\xi}_a \rangle}{\langle \vect{\xi}_a, \vect{\xi}_a \rangle} \nonumber \\
    &= - m_a \Omega  \frac{\int dr_{\rm ns} \varrho r_{\rm ns}^2 \left(2\xi_{a,r}\xi_{a,h} + \xi_{a,h}^2\right)}{\int dr_{\rm ns} \varrho r_{\rm ns}^2 \left[\xi_{a,r}^2 + l_a(l_a+1) \xi_{a,h}^2\right] }, \nonumber \\
    &\simeq - \frac{m_a}{l_a} \Omega.
    \label{eq:Delta_omega_a_corot}
\end{align}
Here $\xi_{a,r}$ and $\xi_{a, h}$ are the radial and tangential components of $\vect{\xi}_{a}$ (see, e.g., Ref. \cite{Weinberg:12}),
which are evaluated for a non-rotating star. 
In the last line we have approximated the density $\varrho$ as a constant and $\xi_{a, r}{\simeq} l\xi_{a, h}{\sim} r_{\rm ns}^{l-1}$ for the f-modes. Note that our result shows good agreement with, e.g., table I in Ref.~\cite{Lai:21} for different polytropic models.   Higher-order spin effects are ignored in this study but see e.g., Refs.~\cite{Dewberry:22, Pnigouras:24}. We also ignore the gravitational redshift, relativistic frame dragging~\cite{Steinhoff:21}, post-Newtonian spin-tidal couplings~\cite{Abdelsalhin:18}, and nonlinear hydrodynamic corrections \cite{Yu:23a}  to the mode frequency in this analysis for simplicity. 

We normalize the modes as 
\begin{equation}
    \omega_{a0} \left(\omega_{a0} + \omega_{b0}\right) \la \vect{\xi_a}, \vect{\xi_b}\ra = E_1 \delta_{ab} = \omega_{a0}\epsilon_a \delta_{ab},
\end{equation}
where $E_1\equiv M_1^2/R_1$. 
We further define~\cite{Weinberg:12} 
\begin{align}
    U_a &= \frac{\la\xi_a, -\nabla U \ra}{E_1} \nonumber \\
    &= W_{lm} I_a \left(\frac{M_2}{M_1}\right) \left(\frac{R_1}{r}\right)^{l+1} e^{-i m_a(\phi - \Omega t)},
\end{align}
where $W_{lm} = 4\pi Y_{lm}(\pi/2, 0)/(2l+1)$ and $I_a$ is the overlap integral given by
\begin{equation}
    I_a = \frac{1}{M R^l}\int d^3 \vect{x} \varrho \vect{\xi}_a^\ast \cdot \nabla (r_{\rm ns}^l Y_{lm}). 
\end{equation}
This allows us to write
\begin{equation}
    \dot{q}_a + i \omega_a q_a = i \omega_{a0} U_a. 
    \label{eq:q_a}
\end{equation}
We adopt the same reality condition as used in Ref.~\cite{Schenk:02} and require that $q_{\omega_a, m_a} = q^\ast_{-\omega_a, -m_a}$ and $\vect{\xi}_{\omega_a, m_a}=\vect{\xi}^\ast_{-\omega_a, -m_a}$. In other words, for two modes with opposite signs of the frequency $\omega_a$ and the azimuthal quantum number $m_a$ (but other quantum numbers are the same), their amplitudes and eigenfunctions are related by complex conjugation~\cite{Schenk:02}. This condition further requires that $I_{m_a} = (-1)^{m_a} I_{-m_a}$ for the overlap integral. 

While the mode expansion should include a spectrum of modes, in an NS the f-mode strongly dominates the tidal coupling (see, e.g., \cite{Andersson:20}). Therefore, in our numerical analysis, we will focus on the $l_a=2$ and $l_a=3$ f-modes and ignore all the other modes. We will nonetheless maintain the generality of our analytical expressions and keep summations over modes. 
The analytical nature of our study means our results apply to a generic choice of NS equation of state (EOS) as long as one chooses the appropriate values of $(R_1, \omega_a, I_a)$ for given $M_1$. When presenting numerical results to validate our analytical expressions, a NS model consistent with the SLy~\cite{Douchin:01} EOS is assumed by default, which has $M_1=1.4\,M_\odot$ and $R_1=11.7\,{\rm km}$~\cite{Read:09}. The companion is treated as a point particle (PP) with $M_2=M_1$. The initial orbit is assumed to be circular. We assume $\omega_{a0}= 2\pi \times 1.8\times10^3\,{\rm Hz}$ ($M_t \omega_{a0}=0.08$) and $I_a=0.18$ for the $l_a=2$ f-modes, and $\omega_{a0}= 2\pi \times 2.5\times10^3\,{\rm Hz}$ ($M_t \omega_{a0}=0.11$) and $I_a=0.11$ for the $l_a=3$ f-modes~\cite{Steinhoff:21}. The overlap integrals $I_a$ are related to the (adiabatic) Love numbers via Eq.~(\ref{eq:k_lm_ad}) and the chosen values lead to $k_{20}=0.08$ and $k_{30}=0.02$.

To have an extended post-resonance region, we consider a rapidly spinning NS with $\Omega = - 2\pi \times 850\, {\rm Hz}$ when presenting numerical results. The minus sign ($\Omega<0$) indicates that the spin is anti-aligned with the orbital angular momentum. This corresponds to a dimensionless spin of $-0.45$. 
While this value is higher than the spin rate of J1748-2446ad \cite{Hessels:06}, it is still less than half of the maximum spin rate a NS can in principle have. For the assumed SLy EOS, the maximum spin rate is estimated to be $|\Omega_{\rm max}| = 2\pi \times 1.8\times10^3\, {\rm Hz}$ \cite{Read:09}. 
A NS with a softer EOS will typically have higher mode frequencies, yet such a NS can also support a higher spin rate because both effects scale as the dynamical frequency of the NS $\sqrt{M_1/R_1^3}$. 
As a caveat, we emphasize that our treatment of the background NS is only accurate to the linear order in $\Omega/\omega_1$ where $\omega_1^2=M_1/R_1^3$. This modifies the corotating frame mode frequencies by the Coriolis force according to Eq.~(\ref{eq:Delta_omega_a_corot}). We do not account for the effect of rotation on other background quantities, and we also ignore rotational corrections to the eigenfunctions.
For the modes to be resonantly excited before the merger, approximated by the location of the innermost stable circular orbit (ISCO) $r_{\rm isco}=6 M_t$ ($\simeq 2.1 R_1$ for the SLy EOS), or a GW frequency of $f_{\rm isco}=\omega_{\rm isco}/\pi =1570\,{\rm Hz}$, the critical spin rate is
\begin{align}
    \Omega_{\rm crit}^{\rm (SLy)} &\simeq \frac{l_a \omega_{\rm isco} - \omega_{a0}}{l_a-1}, \nonumber \\
    &\simeq
    \begin{cases}
        -2\pi \times 260\,{\rm Hz} \text{ for $l_a=2$}, \\
        -2\pi \times 53\,{\rm Hz} \text{ for $l_a=3$}. \\
    \end{cases}
    \label{eq:Omega_crit}
\end{align}
We have used Eq.~(\ref{eq:Delta_omega_a_corot}) and set $m_a=l_a$ for the estimation above.  
On the other hand, for a harder EOS like H4~\cite{Lackey:06}, the binary contacts each other ($r=2R_1$) outside the ISCO. The GW frequency when they contact is about $f_{r=2R_1}=1345$ Hz. In this case, the critical spin rates are 
\begin{align}
    \Omega_{\rm crit}^{\rm (H4)} &\simeq \frac{l_a \omega_{r=2R_1} - \omega_{a0}}{l_a-1}, \nonumber \\
    &\simeq
    \begin{cases}
        -2\pi \times 140\,{\rm Hz} \text{ for $l_a=2$}, \\
        +2\pi \times 5\,{\rm Hz} \text{ for $l_a=3$}. \\
    \end{cases}
\end{align}
In this case, the $l_a\geq 3$ f-modes can be resonantly excited even for a non-spinning NS.

Motivated by the general solution of a driven harmonic oscillator, we solve the mode amplitude evolution analytically by decomposing it as
\begin{align}
    q_a &= q_a^{\rm (eq)} + q_a^{\rm (dyn)} \nonumber \\
    &= b_a^{(\rm eq)} e^{-i m_a (\phi - \Omega t)} + c_a^{\rm (dyn)} e^{- i (\omega_a t - \psi_r)}.
    \label{eq:mode_decomposition}
\end{align}
In other words, we decompose the mode into an equilibrium component that varies with the driving force,
and a dynamical tide component that varies at the oscillator's eigenfrequency. We also introduce slow varying mode amplitudes $b_a^{\rm (eq)}$ and $c_a^{\rm (dyn)}$ with the fast-evolving phase components factored out. We will in the following sections describe closed-form solutions to each component. 

Before proceeding, we emphasize that in our work the different meanings of ``adiabatic'', ``equilibrium'', and ``dynamical''. We will use ``adiabatic'' to mean the zero-frequency limit where the orbital frequency is set to $\omega=\dot{\phi}=0$ in the amplitude equation (while we still keep $r$ and $\phi$). In other words, in the adiabatic limit we set $\dot{q}_a$ to $(i m_a \Omega q_a)$ in Eq.~(\ref{eq:q_a}). The word ``equilibrium'' is used when the mode amplitude can be approximated by Eq.~(\ref{eq:b_a_eq}) [or in many practical cases, Eq.~(\ref{eq:b_a_eq_0})]. Therefore, the equilibrium solution in our language contains the finite-frequency ($\omega>0$) correction to the adiabatic solution. This is different than the equilibrium tide used in, e.g., Refs. \cite{Weinberg:12, Weinberg:16}, for the zero-frequency limit (which is our adiabatic tide). Our convention is also different from that used in, e.g., Refs.~\cite{Steinhoff:16, Hinderer:16, Steinhoff:21}, where the finite-frequency correction was referred to as the dynamical tide. In our notation, however, the dynamical tide will mean strictly the resonantly excited component that oscillates at the mode's eigenfrequency.

\subsection{Equilibrium tide via resummation}
\label{sec:eq_tide}

Let $b_a = q_a e^{i m (\phi - \Omega t)}$ and $V_a = U_a e^{i m (\phi - \Omega t)}$, the equation for $b_a$ is thus
\begin{equation}
    \dot{b}_a + i (\omega_a - m_a \sigma) b_a = i \omega_{a0} V_a, 
    \label{eq:b_a}
\end{equation}
where $\sigma = \omega - \Omega$ and $\omega = \dot{\phi}$. 
Prior to resonance, $b_a$ is slowly varying, allowing us to obtain an approximate equilibrium solution by ignoring the $\dot{b}_a$ term~\cite{Lai:94c},
\begin{equation}
    b_a^{(0)} = \frac{\omega_{a0}}{\omega_a - m_a \sigma} V_a = \frac{\omega_{a0}}{\Delta_a } V_a, 
    \label{eq:b_a_eq_0}
\end{equation}
where $\Delta_a = \omega_a - m_a \sigma = (\omega_a + m_a \Omega) - m_a \omega$. 
Note that while $b_a^{(0)}$ gives a good approximation to $b_a$ when $m_a \sigma < \omega_a$, it diverges when the mode becomes resonant with $\Delta_a =0$ or $m_a \omega = \omega_a + m_a \Omega$. 
However, the divergence is not real as the mode will only resonate with the orbit for a finite amount of time~\cite{Lai:94c}. 


To prevent the divergence in the algebraic solution of the equilibrium tide, we proceed as follows. First, we write $b_a = b_a^{(0)} + b_a^{(1)} + ...$, and obtain the next order approximation to $b_a$ still under the equilibrium (i.e. away from resonance) assumption as 
\begin{align}
    b_a^{(1)} &= \frac{i}{\Delta_a} \dot{b}_a^{(0)}, \nonumber \\
    & = \frac{i b_a^{(0)}}{\Delta_a} 
    \left[\frac{m_a \omega }{\Delta_a}\frac{\dot\omega}{\omega} - (l_a+1) \frac{\dot{r}}{r}\right]. 
\end{align}
Inspired by Ref.~\cite{Lai:94c}, we now resum $b_a$ as 
\begin{align}
    &b_a^{(\rm eq)} = b_a^{(0)} + b_a^{(1)} + ... 
    \simeq \frac{b_a^{(0)}}{1 - b_a^{(1)}/b_a^{(0)}}\nonumber \\
    &=\frac{\Delta_a \omega_{a0} V_a}{
    \Delta_a^2-i\left[m_a\dot{\omega} - \Delta_a (l_a+1)\frac{\dot{r}}{r}\right]}. 
    \label{eq:b_a_eq}
\end{align}
Note that the equilibrium solution $b_a^{(\rm eq)}$ is everywhere finite. The orbital decay acts as an effective damping term. When resonance happens, we have $b_a^{\rm (eq)}=0$. 

\subsection{Dynamical tide}
\label{sec:dyn_tide}
After constructing an algebraic equation of the equilibrium tide $q_a^{(\rm eq)}= b_a^{(\rm eq)} \exp\left[-im_a(\phi - \Omega t)\right]$, we now solve for the dynamical tide $q_a^{\rm (dyn)} = q_a - q_a^{\rm (eq)}$. 
For this, it is convenient to define $c_a^{\rm (dyn)} = q_a^{\rm (dyn)}\exp\left[i(\omega_a t - \psi_{a,r}) \right]$, where $\psi_{a,r} = (\omega_a + m_a \Omega) t_r - m_a \phi_r$. The subscript ``r'' under a quantity means the quantity is evaluated on resonance when $\Delta_a=0$. 
The equation of motion for $c_a^{\rm (dyn)}$ is 
\begin{align}
    &\dot{c}_a^{\rm (dyn)} = i\omega_{a0} V_a^{\rm (dyn)} e^{i\left[ \omega_a t - m_a (\phi - \Omega t) - \psi_{a,r} \right]}, 
\end{align}
with 
\begin{equation}
    i\omega_{a0} V_a^{\rm (dyn)} = i\omega_{a0} V_a - \dot{b}_a^{(\rm eq)} - i \Delta_a b_a^{\rm (eq)}
\end{equation}.

Note that $c_a$ has net accumulations only when $\Delta_a \simeq 0$, as the phase in the exponential is stationary and can be approximated as
\begin{equation}
    \omega_a t - m_a (\phi - \Omega t) - \psi_r \simeq - \frac{m_a}{2} \dot{\omega}_r \tau^2.
\end{equation}
where we have defined a shifted time $\tau = t- t_r$.
The dominant driving term expanded around mode resonance ($\Delta_a=0$) is given by
\begin{align}
    i\omega_{a0} V_{a}^{\rm (dyn)} 
    &\simeq 
    2 i \omega_{a0} V_{a, r} \nonumber \\
    \times& \frac{\left(1+\frac{2l-9}{6}  \tau / t_{\rm gw, r} \right)}{\left[1 -\frac{2}{3}(l+1) \tau / t_{\rm gw, r}+ i m_a \dot{\omega}_{\rm pp, r} \tau^2\right]^2}.
\end{align}
Since we solve the tidal perturbation to the first order, the PP orbit has been adopted to eliminate $\dot{r}, \ddot{r}, \ddot{\omega}$ in terms of $\dot{\omega}_{\rm pp}$, and $t_{\rm gw}=\omega/\dot{\omega}_{\rm pp}$ [see Eq.~(\ref{eq:domega_pp})]. 
Note that the driving term vanishes as $|\tau|$ increases, highlighting the fact that the dynamical tide's excitation is localized near mode resonance in both phase and amplitude.

We can evaluate the dynamical tide amplitude as 
\begin{align}
    c_a^{\rm (dyn)}
    \simeq& \frac{\omega_{a0} V_{a,r}}{\sqrt{s_a \frac{m_a}{2} \dot{\omega}_{\rm pp, r}}} F(u) \nonumber \\
    \simeq& \sqrt{s_a \frac{2}{m_a} \frac{5}{96 \eta}}  W_{lm} I_a \left(\frac{M_2}{M_1}\right) \left(\frac{R_1}{M_t}\right)^{(l+1)} \nonumber \\
    &\times \left(M_t \omega_r\right)^{(4l-7)/6}\left(M_t \omega_{a0}\right) F(u)
    \label{eq:c_a_dyn}
\end{align}
where in the second line we have used the PP orbit to express $V_{a, r}$ in terms of $\omega_r$. The dimensionless, order unity quantity $F(u)$ describes the excitation of the dynamical tide and it reads
\begin{equation}
    F(u) = \int_{-\infty}^u \frac{2i(1+a_1 u )}{\left(1+ a_2 u + 2i s_a u^2\right)^2 } e^{-i s_a u^2} du,
    \label{eq:Fu}
\end{equation}
and 
\begin{align}
    &u = \left(\sqrt{s_a m_a \dot{\omega}_{\rm pp, r}/2}\right) \tau, \label{eq:u_def}\\
    &\tau = t - t_r= \frac{5}{256} \eta M_t\left[(M_t\omega_r)^{-8/3} - (M_t\omega)^{-8/3} \right].\label{eq:tau_vs_omega}
\end{align}
Using Eq.~(\ref{eq:tau_vs_omega}), we effectively treated $u=u(\omega)$ because the instance of resonance (when $\omega = \omega_r$) is most straightforwardly identified using frequency. Here we still use the PP orbit to convert time and frequency, which is accurate to the linear order in $(\Delta r/r)$ but loses accuracy when $(\Delta r/r)^2$ corrections become important. Such nonlinear corrections are left for future studies. 
Further, $s_a={\rm Sign}[m_a]$. The constants $a_1 = [(2l-9)/6] \sqrt{ (2s_a/m_a) / (\omega_r t_{\rm gw, r})}$ and $a_2=[-2(l+1)/3]\sqrt{ (2s_a/m_a) / (\omega_r t_{\rm gw, r})}$ are both small quantities because $(\omega_r t_{\rm gw, r}) \gg 1$.

\begin{figure}
    \centering
    \includegraphics[width=0.95\linewidth]{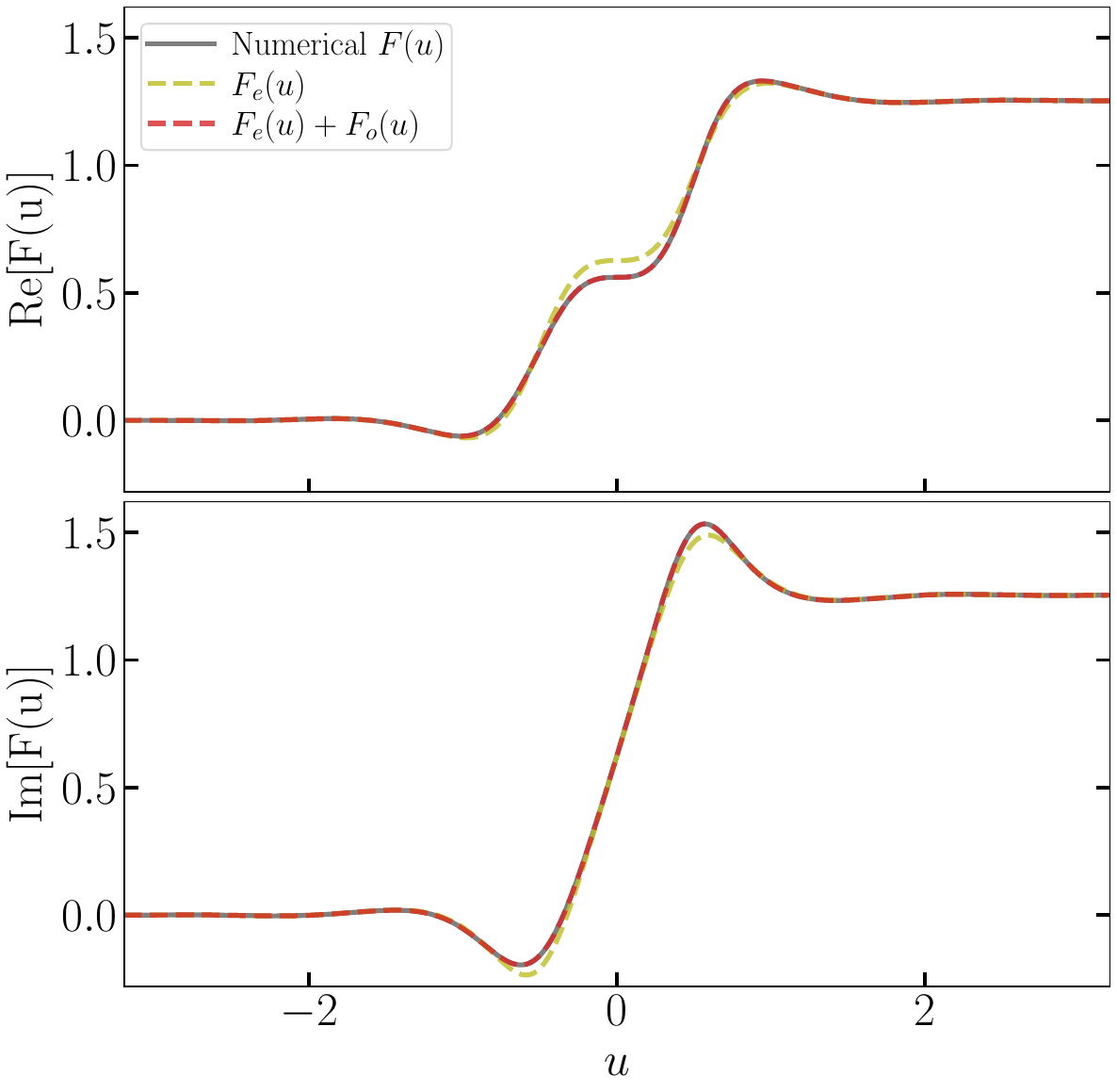}
    \caption{Real and imaginary parts of $F(u)$. The $F_e(u)$ part (whose integrand is even with respect to $u$; red-dashed lines) dominates the result and it is only a function of $u$. The $F_o(u)$ component (included in the olive-dashed lines) corrects the solution near mode resonance and vanishes when $u\to \pm\infty$. It depends on an overall scaling factor $a_3\propto (\omega t_{\rm gw})^{-1/2}$. In the example shown, $a_3=-0.18$. }
    \label{fig:Fu}
\end{figure}

While $F(u)$ can be numerically integrated easily, it is nonetheless instructive to derive its analytical expressions to know the asymptotic behaviors. For this, we decompose the integrand excluding the phase into two components that are respectively even and odd about $u=0$, with
\begin{equation}
    f_e(u) = \frac{2i }{\left(1+2i s_a u^2\right)^2}, \ \  
    f_o(u) = \frac{- 4i a_3 u }{\left(1 + 2i s_a u^2\right)^3}, 
\end{equation}
where $a_3 {=} a_2 {-} a_1/2{=}{-}[(10l{-}1)/12] \sqrt{ (2s_a/m_a) / (t_{\rm gw, r}\omega_r)}$. Since $a_3 \propto (\omega_r t_{\rm gw, r})^{-1/2}\ll 1$, we only keep $a_3$ to the linear order. Note that only the even component has net accumulation across mode resonance. The odd component is a small correction important only near resonance and its significance decreases for modes excited earlier in the inspiral. 
The two components can be separately integrated, leading to 
\begin{widetext}
\begin{align}
    F_e(u) = &\int_{-\infty}^{u} f_e(u) e^{-i s_a u^2} du,\nonumber \\
    = \left(4+16 u^4\right)^{-1}
    &\left\{s_a \left[4u\sin(u^2) + 8u^3\cos(u^2) + \sqrt{2\pi} (1+4u^4)\left(1 + 2{\rm Fs}\left[\sqrt{\frac{2}{\pi}}u\right] \right) \right] \right. \nonumber \\
    &\left. \ \  + i\left[4u\cos(u^2) - 8u^3\sin(u^2) + \sqrt{2\pi} (1+4u^4)\left( 1 + 2{\rm Fc}\left[\sqrt{\frac{2}{\pi}}u\right] \right) \right]\right\}.
    \label{eq:Fe}
    \\
    F_o(u) = &\int_{-\infty}^{u} f_o(u) e^{-i s_a u^2} du,\nonumber \\
    = &\frac{a_3}{8}
    \left\{
    -s_a\sqrt{e} {\rm Ei}\left[-\frac{1+2i s_a u^2 }{2}\right] + 2 i s_a e^{-i s_a u^2} \frac{1-2i s_a u^2}{(1 + 2i s_a u^2)^2} - i \sqrt{e} \pi 
    \right\} \nonumber \\
    \simeq &\frac{0.3654 s_a a_3}{(1+1.3685 i s_a u^2)^4}. 
    \label{eq:Fo}
\end{align}
\end{widetext}
and $F(u) = F_e(u) + F_o(u)$. In the function above, ${\rm Fs}$ and ${\rm Fc}$ are Fresnel integrals, and ${\rm Ei}$ is the exponential integral. Asymptotically, we have $F_e(-\infty) = F_o(\pm \infty)=0$ and $F_e(+\infty) = (\sqrt{\pi/2}) (i + s_a)\simeq 1.25(i + s_a)$. At mode resonance, $F_e(0) = F_e(\infty)/2=(\sqrt{\pi/8}) (i + s_a)$ and $F_o(0) = s_a a_3 (2 -\sqrt{e}{\rm Ei}[-1/2] )/8\simeq 0.3654 s_a a_3$. Fig.~\ref{fig:Fu} shows the real and imaginary parts of $F(u)$ computed from the sum of $F_e(u)$ and $F_o(u)$ in Eqs. (\ref{eq:Fe}) and (\ref{eq:Fo}) and from numerical integration of Eq.~(\ref{eq:Fu}).

Going back to $b_a^{\rm (dyn)}$ with
\begin{equation}
    b_a^{\rm (dyn)} = c_a^{(\rm dyn)} e^{i\left[m_a(\phi-\phi_r) - (\omega_a + m_a \Omega)\tau\right]}, 
    \label{eq:c_a_2_b_a}
\end{equation}
where we use the leading-order quadrupole formula to write $\phi-\phi_r$ as functions of frequencies
\begin{align}
    \left(\phi-\phi_r\right)_{\rm pp} 
    =\frac{1}{32} \frac{M_t}{\mu} \left[(M \omega_r)^{-5/3} - (M \omega)^{-5/3} \right].
    \label{eq:phi_vs_omega_pp}
\end{align}

\begin{figure}
    \centering
    \includegraphics[width=0.95\linewidth]{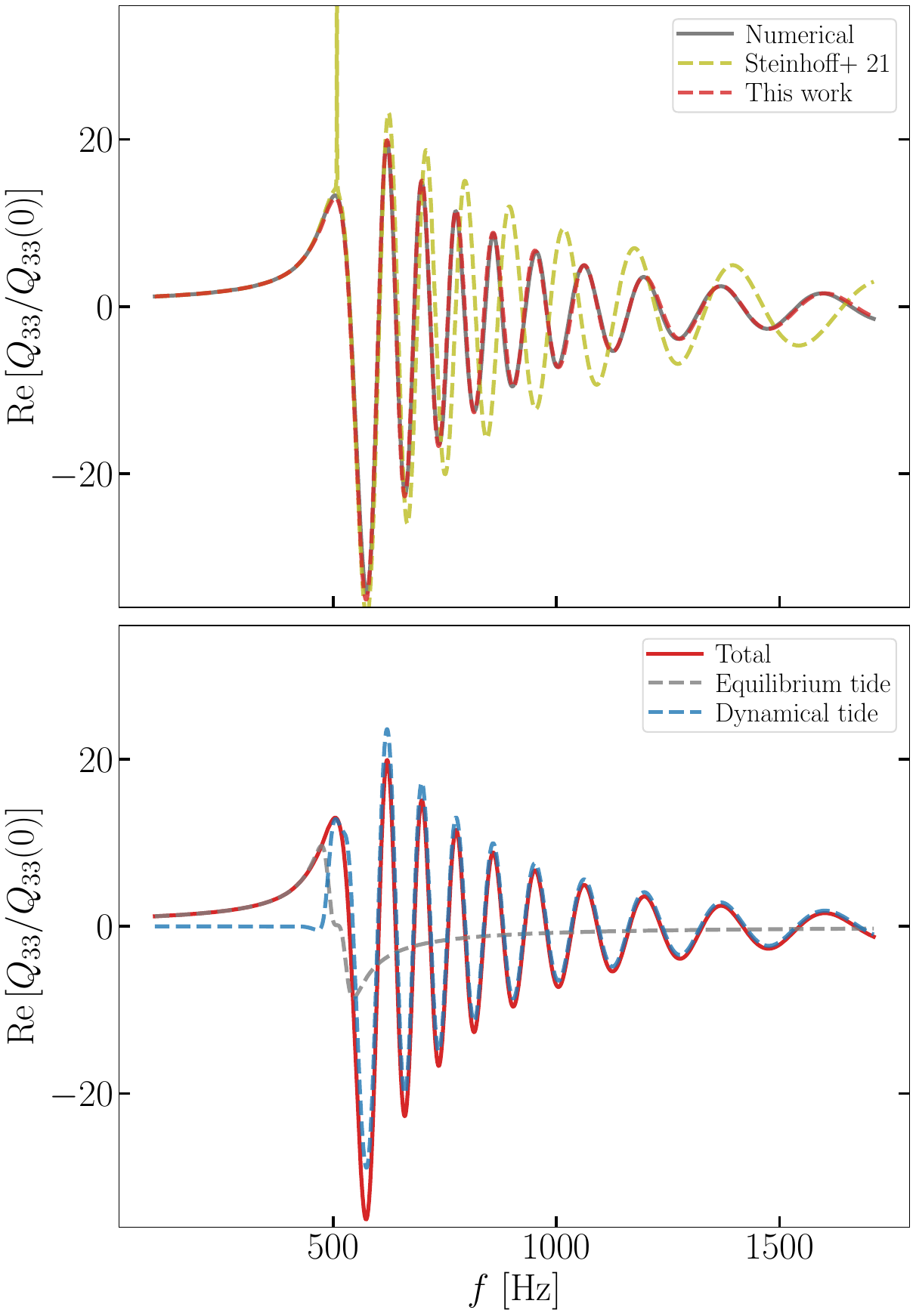}
    \caption{Top: Comparison of different analytical approximations for the $l=m=3$ mass multipole. We have normalized the results by the adiabatic limit, and the result also corresponds to the Love number as $k_{33}/k_{33}^{\rm (eq)}(0) = Q_{33}/Q_{33}(0)$, see Eq. (\ref{eq:k_lm}). The solution from Ref. \cite{Steinhoff:21} (olive-dashed line) is corrected according to Footnote~\ref{ft:vs_S+21} so that it applies to $l=3$. Because the S+21 solution requires the subtraction of two diverging terms at mode resonance (around 510 Hz), it may cause numerical artifacts. Furthermore, the phase and amplitude of the multipole are inaccurate in the post-resonance regime. The new solution (red-dashed line) however fixes these issues and matches the numerical solution throughout the entire evolution.
    Bottom: Decomposition of the solution into an equilibrium tide component and a dynamical tide component. The equilibrium component vanishes smoothly at mode resonance, facilitating the numerical implementation. The dynamical tide is excited at mode resonance and dominates the post-resonance solution. 
    In this figure and all other figures for the $l=3$ tide, we include the tidal back-reaction on the orbit only from the $l=3$ tide and exclude the $l=2$ tide. }
    \label{fig:Q33}
\end{figure}

The comparisons of the new analytical solution with numerical values and with previous results from Ref. \cite{Steinhoff:21} are presented in Fig.~\ref{fig:Q33} for the mass multipole $Q_{33}$ with $l=m=3$. The relation between mass multipole and mode amplitude is given later in Eq.~(\ref{eq:multipole_from_amp}). We have normalized the result by its adiabatic limit obtained by replacing $b_a$ in Eq. (\ref{eq:multipole_from_amp}) with $\omega_{a0}V_a/(\omega_{a}+ m_a \Omega_s)$. 
The gray line in the top panel is the numerical result from solving the coupled differential equations governing the evolution of the mode and the orbit [see Eqs.~(\ref{eq:ddr}) and (\ref{eq:ddphi})]. In this process, we keep the tidal back-reaction due to the $l=3$ tide but drop the back-reactions from the $l=2$ tide. The $l=2$ tide can cause the orbit to deviate significantly from the PP limit, requiring higher-order corrections that are beyond the scope of this work (also Ref.~\cite{Steinhoff:21}) that solves the tide at the linear order (in both back-reaction and fluid displacement; see later the discussion about Fig~\ref{fig:dphi_l2} and Sec. \ref{sec:conclusion}). Our new solution agrees well with the numerical one throughout the inspiral as shown in the red-dashed line in the top panel, and its decomposition into the equilibrium tide [Eq. (\ref{eq:b_a_eq})] and the dynamical tide [Eqs. (\ref{eq:c_a_dyn}) and (\ref{eq:c_a_2_b_a})] is presented in the bottom panel. In comparison, the solution constructed following Ref.~\cite{Steinhoff:21} is given in the olive-dashed line.\footnote{In particular, we use the part after $b_l$ in eq. (6.10) from Ref.~\cite{Steinhoff:21}. The relation between the Love number and mass multipole is given by Eq.~(\ref{eq:k_lm}). Note that their $\Delta \omega_{0l}$ measures the frequency shift of a mode in the inertial frame, so it corresponds to our $\Delta \omega_a + m_a\Omega$. } While it has excellent accuracy in the pre-resonance regime, it shows a spike at 510 Hz that is due to the imperfect numerical cancellation between the diverging term in Eq.~(\ref{eq:b_a_eq_0}) and the counter term [second line in their eq. (6.10)] at mode resonance. When extended to the post-resonance regime, the solution from Ref.~\cite{Steinhoff:21} loses accuracy in both amplitude and phase, which we will discuss quantitatively in the subsequent section (Sec.~\ref{sec:mode_amp_vs_old}). 

As an aside, we also prove using Eq.~(\ref{eq:c_a_dyn}) that the dynamical tide is a converging series as $l$ increases. For this, we note first that $\omega_r\simeq \omega_{a0}/l + [(l-1)/l] \Omega$ for the f-modes that are most likely to be excited with $m_a\simeq l$. Further, 
\begin{equation}
    M_t \omega_{a0} \propto l^{1/2} \left(\frac{M_1}{M_t}\right)^{1/2}\left(\frac{R_1}{M_t}\right)^{-3/2}. 
\end{equation}
In the slow-rotating limit with $|\Omega|\ll |\omega_{a0}|$, we have
\begin{equation}
    c_a^{(\rm dyn)} \propto \left(\frac{1}{l}\right)^{\frac{4l-13}{12}} \left(\frac{M_1}{M_t}\right)^{\frac{4l-1}{12}} \left(\frac{R_1}{M_t}\right)^{\frac{5}{4}}. 
\end{equation}
This is clearly a converging series. In the case where $|\Omega|$ dominates, we first note that only when $\Omega>0$ can there be infinitely many f-modes be resonantly excited all at $\omega_r\simeq \Omega$. The maximum $\Omega$ a NS can have is $(M_1/R_1^3)^{(1/2)}$, which leads to 
\begin{equation}
    c_a^{(\rm dyn)} \propto l^{1/2} \left(\frac{M_1}{M_t}\right)^{\frac{4l-1}{12}} \left(\frac{R_1}{M_t}\right)^{\frac{5}{4}}. 
\end{equation}
In the high-$l$ limit, the $(M_1/M_t)<1$ term decreases exponentially while the $l^{1/2}$ increases only as polynomial. Therefore, the series is also converging. For $\Omega<0$, only modes with $\omega_{a0}>-(l-1)\Omega$ can experience resonance. Because this is a finite set of modes, divergence cannot happen.

\subsection{Comparison with previous analysis}
\label{sec:mode_amp_vs_old}

We elaborate further on the relation between our new solution and that obtained in previous analyses such as Ref.~\cite{Steinhoff:16} and \cite{Steinhoff:21}.  

We note that Eq.~(\ref{eq:b_a}) can be solved locally around mode resonance by expanding 
\begin{equation}
    \omega_a - m_a\sigma \simeq m_a \dot{\omega}_r \tau \simeq m_a \frac{\dot{\omega}_r}{\omega_r} \varpi, 
\end{equation}
where $\varpi=(\phi-\phi_r)$. This leads to the solutions in the resonance region obtained in \cite{Steinhoff:16} and \cite{Steinhoff:21} (in particular, the last line in eq. (6.10) of Ref. \cite{Steinhoff:21}). In this case, we have
\begin{align}
    b_a^{\rm (loc)} &\simeq i \omega_{a0} V_{a,r} e^{i \frac{m_a}{2} \dot{\omega}_r t^2} \int_{-\infty}^t e^{-i \frac{m_a}{2} \dot{\omega}_r t^2} dt,\nonumber \\
    &\simeq i \frac{\omega_{a0} V_{a,r}}{\omega_r} e^{i \frac{m_a}{2} \frac{\dot{\omega}_r}{\omega_r^2} \varpi^2}\int_{-\infty}^\varpi e^{-i \frac{m_a}{2} \frac{\dot{\omega}_r}{\omega_r^2} \varpi^2} d\varpi,\nonumber \\
    &= \frac{\omega_{a0} V_{a,r}}{\sqrt{s_a \frac{m_a}{2} \dot{\omega}_r}} G(v)
\end{align}
where\footnote{Note that $\hat{t}$ in Ref~\cite{Steinhoff:21} is related to our $v$ as $\hat{t}= \sqrt{(2/m_a)(8/3)} v$ and their $\epsilon_l=(3/8)(\dot{\omega}_r/\omega_r^2)$. Our analysis also indicates that a factor of $\sqrt{m_a/2}=\sqrt{l_a/2}$ in the phases in eq. (6.11) of Ref. \cite{Steinhoff:21} is missing. The magnitude of the last term in their eq. (6.10) lacks a factor of $\sqrt{l_a/2}$. \label{ft:vs_S+21}} 
$v=\left(\sqrt{s_a m_a \dot{\omega}_{\rm pp, r}/2}/\omega_r\right) \varpi =\left(\omega/\omega_r\right) u$ and
\begin{align}
    G(v) &= i e^{i s_a v^2}\int_{-\infty}^{v} e^{-i s_a v^2} dv \nonumber \\
    &=\sqrt{\frac{\pi}{8}} (s_a + i)e^{is_a v^2} \nonumber \\
    \times &
     \left\{1 + (1+i s_a){\rm Fc}\left[\sqrt{\frac{2}{\pi}}v\right] 
             + (1-i s_a){\rm Fs}\left[\sqrt{\frac{2}{\pi}}v\right] 
    \right\}.
\end{align}
Note that $G(0)=F_e(0)$ and $\lim_{v\to +\infty} G(v) e^{-i s_a v^2}=F(+\infty)$, indicating the consistency in both the mode amplitude at resonance and their asymptotic values. However, the phase of $b_{a}^{\rm (loc)}$ is accurate only in the vicinity of mode resonance. It also overestimates the magnitude of oscillations in the post-resonance region.  
Another way to obtain a local solution is to solve $c_a$ [cf. Eq.~(\ref{eq:c_a_dyn}) without the ``(dyn)'' superscript] using a stationary phase approximation, and then transfer the resultant $c_a$ to $b_a$ using Eq.~(\ref{eq:c_a_2_b_a})~\cite{Lai:94a}. Such a solution gives the correct phase in the post-resonance regime but not in the prior-resonance regime. Indeed, the general solution of a driven oscillator should contain two terms varying at distinct frequencies as in Eq.~(\ref{eq:mode_decomposition}), and a local solution with a single term can thus be extended accurately only in one direction. The discussion therefore highlights the necessity of decomposing the tide into the equilibrium and dynamical components in the presence of mode resonance.

\section{Back-reaction on the orbit using effective Love number}
\label{sec:love_num}


In this section, we first convert the mode amplitudes found in the previous section to mass multipoles of the NS, and then use the mass multipoles to define effective Love numbers following the approach used in Ref.~\cite{Hinderer:16}. The tidal back reactions will then be written in terms of the effective Love number so that they take the same form as in the adiabatic limit, except for the Love number replaced by their frequency-dependent effective values. We highlight that a pure real effective Love number as introduced in Ref. \cite{Hinderer:16} is insufficient to model the dynamics. In particular, it captures only the tidal acceleration in the radial direction. To capture the acceleration in the tangential direction (i.e., the tidal torque), a new dressing factor, originating from the imaginary part of the effective Love number for each $(l, m)$ harmonic, will be introduced. We will demonstrate explicitly that it is this torque that dominates the impact on the orbit near mode resonance.  

We start by relating the multipole moment  of the NS \emph{in the inertial frame} to the amplitude of the associated mode with the same $(l, m)$ as~\cite{Yu:23a},
\begin{align}
    Q_{lm} = M R^l \sum_a^{l_a, m_a} I_a b_a e^{-i m_a \phi},
    \label{eq:multipole_from_amp}
\end{align}
where we have used $\sum_a^{l_a, m_a}$ as a short-hand notation of $\sum_a^{(l_a, m_a)=(l, m)}$. Note further that the summation runs over modes with both signs of eigenfrequencies.
This corresponds to a symmetric trace-free tensor (indicated by the brackets around the indices) in a Cartesian coordinate, 
\begin{align}
    &Q^{\la i_1 ... i_l \ra} = N_l \sum_m {\mathcal{Y}_{lm}^{\la i_1 ... i_l \ra}}^\ast Q_{lm}\nonumber \\
    &=N_l M_1 R^l \sum_m {\mathcal{Y}_{lm}^{\la i_1 ... i_l \ra}}^\ast \sum_a^{l_a, m_a} I_a b_a e^{-i m_a \phi}.
\end{align}
where $N_l = 4\pi l!/(2l+1)!!$ and the tensor ${\mathcal{Y}_{lm}^{\la i_1 ... i_l \ra}}$ is defined in Ref.~\cite{Poisson:14}.
At the same time, the tidal moment (also symmetric and trace-free) can be written as 
\begin{align}
    &\mathcal{E}_{ i_1 ... i_l } \equiv - M_2 \partial_{i_1 ... i_l} \frac{1}{r} \nonumber \\
    &= - l! \frac{M_2}{r^{l+1}}\sum_m W_{lm} e^{-im \phi} {\mathcal{Y}_{lm}^{\la i_1 ... i_l \ra}}^\ast
\end{align}
We can define an effective Love number of each $(l, m)$ harmonic so that it maintains the same form as in the adiabatic limit as
\begin{equation}
    Q_{\la i_1 ... i_l \ra} \mathcal{Y}_{lm}^{\la i_1 ... i_l \ra} =-\frac{2k_{lm}}{(2l-1)!!}R^{2l+1}\mathcal{E}_{ i_1 ... i_l }\mathcal{Y}_{lm}^{\la i_1 ... i_l \ra},
\end{equation}
which leads to 
\begin{align}
    &k_{lm} = \frac{2\pi}{2l+1} \sum_a^{l_a, m_a} I_a^2 \left(\frac{b_a}{V_a}\right) = k_{l, -m}^\ast \in \mathbb{C}.
    \label{eq:k_lm}
\end{align}
The last equality follows the reality condition on the mode amplitude. 
Note that in the non-spinning adiabatic limit,
\begin{equation}
    k_{lm}|_{\omega=\Omega=0} = k_{l0}= \frac{4 \pi}{2l+1} I_a^2.
    \label{eq:k_lm_ad}
\end{equation}
and the Love number is the same for all values of $m$. 
Using the leading-order equilibrium solution from Eq. (\ref{eq:b_a_eq_0}), we have 
\begin{align}
    k_{lm}^{\rm (eq,0)}(\omega)&=k_{l,-m}^{\rm (eq,0)}(\omega)=\frac{\omega_{a0}^2}{\omega_{a0}^2 - (m \sigma - \Delta \omega_a)^2} k_{l0},
\end{align}
where mode $a$ has $(l_a, m_a)=(l, m)$, and we have used the fact that $\Delta \omega_a \propto m_a$ [Eq.~(\ref{eq:Delta_omega_a_corot})], so $\Delta_{m_a} = - \Delta_{-m_a}$. This result agrees with Ref.~\cite{Lai:21}.  
Because of rotation, we have 
\begin{equation}
    k_{lm}^{\rm (eq,0)}(\omega=0) = \frac{\omega_{a0}^2}{\omega_{a0}^2 - (m \Omega+ \Delta \omega_a)^2} k_{l0}.
    \label{eq:k_lm_eq_omega_0}
\end{equation}
Note that the difference between $k_{lm}^{\rm (eq, 0)}$ and $k_{l0}$ are due to the finite frequency response of a mode as in Eq.~(\ref{eq:b_a_eq_0}). Their difference is on the order $(\Omega/\omega_{a0})^2$. As a caveat, the mode structure can be modified by rotation at order $(\Omega/\omega_{a0})^2$ (e.g., due to the centrifugal force), which is not accounted for in our current framework as Eq.~(\ref{eq:xi_eom}) is only to linear order in rotation. But see Ref.~\cite{Pnigouras:24} for more discussions on rotational corrections beyond Eq.~(\ref{eq:Delta_omega_a_corot}).



We can further define an effective dressing factor $\kappa_l$ for each harmonic $l$ as
\begin{align}
    \kappa_l &= - \frac{(2l-1)!!}{2 k_{l0} R^{2l+1}}\frac{Q_{\la i_1 ... i_l \ra} \mathcal{E}^{i_1 ... i_l}}{\mathcal{E}_{i_1 ... i_l}{\mathcal{E}^{i_1 ... i_l}}} \nonumber \\
    &=\frac{(2l+1)}{4 \pi} \sum_m W_{lm}^2 \frac{k_{lm}}{k_{l0}} \in \mathbb{R},
    \label{eq:kappa_l}
\end{align}
so that $\kappa_l k_{l0}$ becomes the effective Love number in Ref.~\cite{Hinderer:16}. 
Note that $\sum_m W_{lm}^2=4 \pi/(2l+1)$, so when $\omega=\Omega=0$, $\kappa_l=1$. The reality condition that $k_{lm}=k_{l,-m}^\ast$ further requires that $\kappa_l$ is real. 

As we will see shortly, the effective Love number $\kappa_l k_{l0}$ can only capture the radial acceleration of the tidal back-reaction. We hence introduce an additional factor $\gamma_l$ to describe the torque, which we define as 
\begin{equation}
    \gamma_l = \frac{2l+1}{4\pi}\sum_m m W_{lm}^2 {\rm Im}\left[ \frac{k_{lm}}{k_{l0}} \right]. 
    \label{eq:gamma_l}
\end{equation}
Note that $\gamma_l$ vanishes in the equilibrium limit but becomes finite when $k_{lm}$ has imaginary component as $m {\rm Im}[k_{lm}] = -m {\rm Im}[k_{l, -m}]$. 

To derive the conservative part of the dynamics, we write the Hamiltonian of the system to linear order in the NS's rotation~\cite{Schenk:02, Flanagan:07, Yu:23a} as,
\begin{equation}
    H = H_{\rm pp} + H_{\rm mode} + H_{\rm int},
    \label{eq:Hamiltonian}
\end{equation}
where 
\begin{equation}
    H_{\rm pp} = p_r^2/(2\mu) + p_\phi^2/(2\mu r^2) - \mu M_t/r,
\end{equation}
with $p_r = \mu \dot{r}$ and $p_\phi = \mu r^2 \dot{\phi}$, 
\begin{align}
    H_{\rm mode} &= \sum_a^{\omega_a>0} \epsilon_a \omega_a q_a^\ast q_a = E_1\sum_a^{\omega_a>0} \frac{\omega_a}{\omega_{a0}} b_a^\ast b_a,   
\end{align} 
and 
\begin{equation}
    H_{\rm int} = \sum_a^{\omega_a>0}\epsilon_a  \left(i f_a q_a^\ast - i f_a^\ast q_a\right) = -E_1 \sum_a^{\omega_a>0} 
    \left( V_a b_a^\ast + V_a^\ast b_a
    \right).
\end{equation}
The three pieces respectively represent the PP orbit, the eigenmodes inside the NS, and the mode-orbit interaction. While we use $b_a$ and $V_a$ in the above expressions, we emphasize that the generalized displacements we use are $(r, \phi, q_a)$ and their associated momenta are $[p_r, p_\phi,  (i E_1/\omega_{a0}) q_a^\ast]$. The Hamiltonian depends explicitly on time because $V_a b_a^\ast \sim \exp(i m_a \Omega t)$ in the interaction. We will come back to this point later when discussing the energy conservation of the system [Eq.~(\ref{eq:dEdt_mode_phy})].

The dynamics of the orbit is thus given by 
\begin{align}
    &\ddot{r} - r \dot{\phi}^2 + \frac{M_t}{r^2} = g_r, \label{eq:ddr}\\
    &r\ddot{\phi} + 2 \dot{r}\dot{\phi} = g_\phi \label{eq:ddphi}
\end{align}
Where $g_r$ and $g_\phi$ are accelerations along the radial and tangential directions and they include both a dissipative contribution due to GW radiation and a conservative piece due to tidal interaction, which we can derive from the interaction Hamiltonian as 
\begin{align}
    &g_{r}^{(t)} = -\frac{1}{\mu}\frac{\partial H_{\rm int}}{\partial r} = -\frac{2E_1}{\mu r} \sum_a^{\omega_a>0} (l_a+1) {\rm Re}\left(V_a b_a^\ast\right) \nonumber \\
    &=-\frac{M_t}{r^2}\frac{M_2}{M_1}\sum_l \frac{(l+1)(2l+1)}{2\pi}\left(\frac{R}{r}\right)^{2l+1}
    \sum_m W_{lm}^2 {\rm Re}\left[k_{lm}\right], \nonumber \\
    &=-2 \frac{M_t}{r^2}\frac{M_2}{M_1}\sum_l (l+1) \kappa_l k_{l0} \left(\frac{R}{r}\right)^{2l+1},
    \label{eq:g_r}
    \\
    &g_{\phi}^{(t)} = -\frac{1}{\mu r} \frac{\partial H_{\rm int}}{\partial \phi}
    =\frac{2E_1}{\mu r} \sum_a^{\omega_a >0}  m_a {\rm Im}\left[ V_a b_a^\ast \right], \nonumber \\
    &=- \frac{M_t}{r^2} \frac{M_2}{M_1}\sum_l \frac{(2l+1)}{2\pi} \left(\frac{R}{r}\right)^{2l+1} \sum_m m W_{lm}^2 {\rm Im}\left[k_{lm}\right], \nonumber \\
    &=-2 \frac{M_t}{r^2}\frac{M_2}{M_1}\sum_l (l+1) \gamma_l k_{l0} \left(\frac{R}{r}\right)^{2l+1},
    \label{eq:g_phi}
\end{align}
where we have used Eqs.~(\ref{eq:k_lm}), (\ref{eq:kappa_l}), and (\ref{eq:gamma_l}) and their (real and imaginary) dressing factors.

Note that we can further write the interaction energy as 
\begin{equation}
    H_{\rm int} = -2 \frac{M_2^2}{r}\sum_l \kappa_l k_{l0} \left(\frac{R}{r}\right)^{2l+1}.
    \label{eq:H_int}
\end{equation}
In the limit where $\Omega\to 0$ and $\omega\to 0$ (i.e., $\kappa_l=1$), we have 
\begin{equation}
    H_{\rm mode}(\omega=\Omega=0) = -\frac{1}{2} H_{\rm int} = \frac{M_2^2}{r}\sum_l k_{l0} \left(\frac{R}{r}\right)^{2l+1}.
\end{equation}
However, in general, there is no simple relation between $H_{\rm mode}$ and $\kappa_l k_{l0}$. 
Note further that when $\Omega\neq0$, $H_{\rm mode}$ is the mode energy in the corotating frame. For a specific mode $a$ in the inertial frame, its canonical energy is~\cite{Friedman:78}
\begin{align}
    &E_{{\rm mode},a}^{\rm (inertial)} = H_{{\rm mode}, a} + \Omega J_{{\rm mode}, a}  \nonumber \\
    =& \frac{\omega_a + m_a \Omega}{\omega_a} H_{{\rm mode},a} = \frac{E_1}{2}\frac{\omega_a+m_a \Omega}{\omega_{a0}} b_a^\ast b_a ,
    \label{eq:E_mode_phy}
\end{align}
where $J_{{\rm mode}, a} = m_a H_{\rm mode}/\omega_a$ is the angular momentum of the mode. Using the asymptotic values of $F_{e}(u)$, we have when $u \to \infty$ (after summing over a mode $a$ and its complex conjugate)
\begin{align}
    &\frac{E_{{\rm mode}}^{\rm (inertial)}(\infty)}{E_1} \simeq \frac{2\pi}{s_a m_a}\frac{\omega_{a0}(\omega_a + m_a \Omega)}{\dot{\omega}_{\rm pp, r}} V_{a, r}^2. \nonumber \\
    &=\frac{5(2l+1)}{192 \eta} W_{lm}^2 k_{l0} \left(\frac{M_2}{M_1}\right)^2 \nonumber \\
    &\times \left(\frac{R_1}{M_t}\right)^{2(l+1)} \left(M_t \omega_{a0}\right) \left(M_t \omega_r \right)^{4(l-1)/3}.
    \label{eq:E_mode_inf}
\end{align}

While $H_{\rm mode}$ does not relate to $\kappa_l$ and $\gamma_l$ in a simple manner, we note that its derivative is 
\begin{align}
    \dot{E}_{\rm mode}^{\rm (inertial)} &= \dot{H}_{\rm mode} + \frac{\partial H_{\rm int}}{\partial t} \nonumber \\
    & = -2 E_1 \sum_a^{\omega_a>0}\left(\omega_a + m_a \Omega\right) {\rm Im}\left[V_a b_a^\ast\right]. 
    \label{eq:dEdt_mode_phy}
\end{align}
We remind readers that the $\partial H_{\rm int}/\partial t$ term originates from the fact that the Hamiltonian depends on time explicitly. 
Together with the total time derivative of $H_{\rm int}$, 
\begin{align}
    &\dot{H}_{\rm int} =2E_1 \sum_a^{\rm \omega_a>0}\left\{\Delta_a {\rm Im} \left[V_a b_a^\ast\right] + \frac{\dot{r}}{r}(l_a + 1){\rm Re}\left[V_a b_a\right]\right\}, 
\end{align}
we have
\begin{equation}
    \dot{E}_{\rm tide} \equiv \dot{E}_{\rm mode}^{(\rm inertial)} + \dot{H}_{\rm int} = - \mu \dot{r} g_r^{(t)} - \omega \mu r g_\phi^{(t)},
    \label{eq:dEtide_dt}
\end{equation}
which describes the expected energy transportation from the orbit to the tide. 

\begin{figure}
    \centering
    \includegraphics[width=0.95\linewidth]{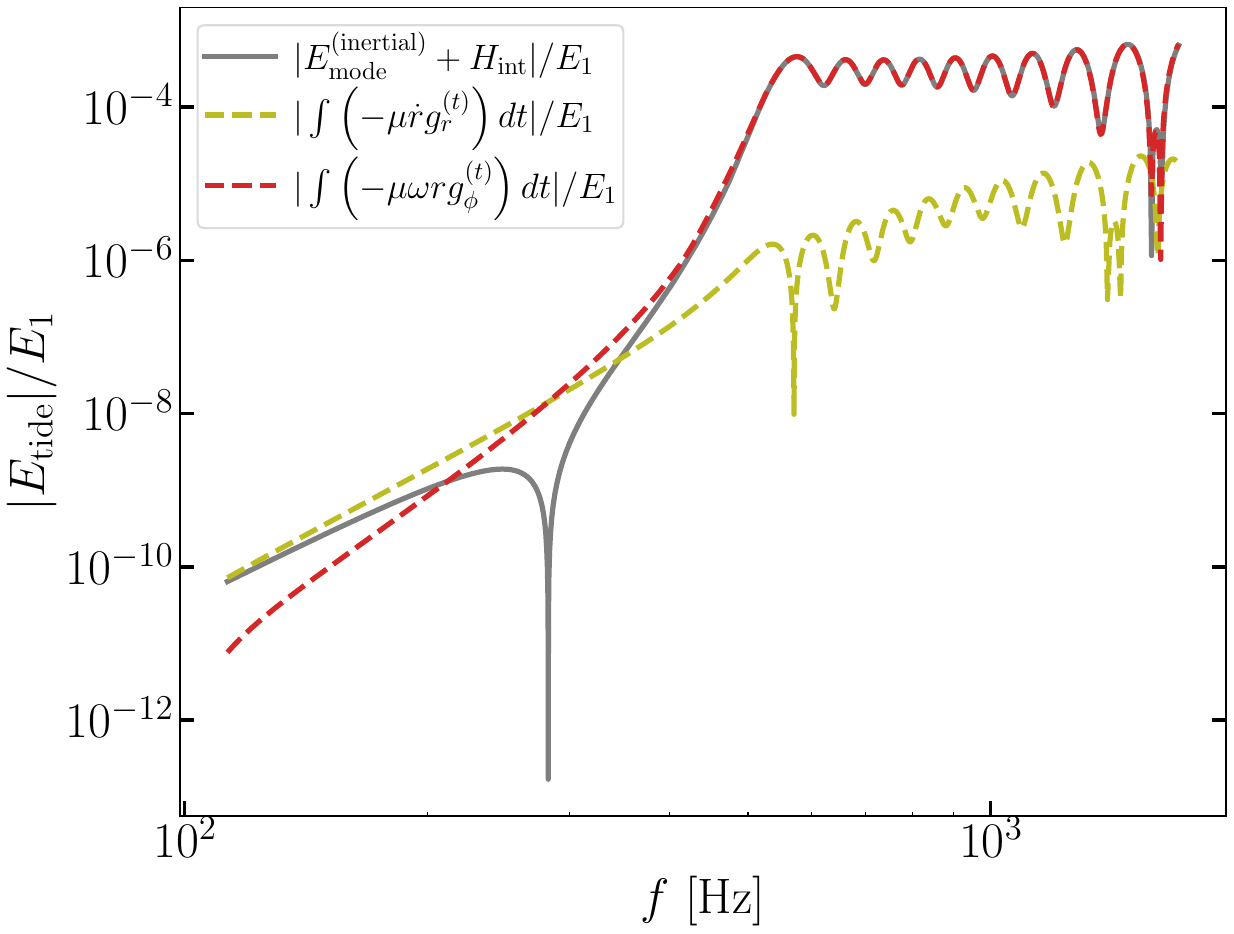}
    \caption{Energy in the $l=3$ tide, $E_{\rm tide} = E_{\rm mode}^{\rm (inertial)}$ and $H_{\rm int}$, shown in gray line. Also shown in the dashed lines are the integrals of the two terms in Eq.~(\ref{eq:dEtide_dt}). In the low-frequency, equilibrium limit, the radial term (olive line) dominates the change and it is negative (the change of the total equilibrium energy is still positive when including $\Delta E_{\rm pp}$). Near mode resonance, the tangential torque (red line; positive) dominates the change of tidal energy. }
    \label{fig:Etide}
\end{figure}

Fig.~\ref{fig:Etide} shows the integral of the Eq.~(\ref{eq:dEtide_dt}). 
In the equilibrium limit with $|\omega_a| \gg \omega\ {\rm and}\ |\omega_a| \gg |\Omega|$, $|{\rm Im}(b_a)/{\rm Re}(b_a)| \sim |\dot{r}/(r \Delta_a)|$, leading to
\begin{equation}
    \frac{\omega r g_\phi^{(t)}}{\dot{r} g_r^{(t)}} \sim \frac{\omega}{\Delta_a} \simeq \frac{\omega}{\omega_a}\ll 1 \text{ (for equilibrium tide)}.
    \label{eq:gphi_vs_gr_eq}
\end{equation}
Therefore, in the equilibrium limit, the torque term (or $\gamma_l$) can be ignored. 
However, when dynamical tide becomes important and $|{\rm Im}(b_a)/{\rm Re}(b_a)| \sim 1$, we have 
\begin{equation}
    \frac{\omega r g_\phi^{(t)}}{\dot{r} g_r^{(t)}} \sim \frac{\omega}{\dot{r}/r}\gg 1\text{ (for dynamical tide)}. 
    \label{eq:gphi_vs_gr_dyn}
\end{equation}
Furthermore, the torque term has a secular contribution when integrated across mode resonance, 
\begin{align}
    & \int_{-\infty}^{\infty} -\omega \mu r g_\phi^{(t)} dt \nonumber \\
    \simeq&  2 E_1 m_a \omega_r V_{a, r} {\rm Im}\left[c_{a,r}^{\rm (dyn)}\int e^{i \frac{m_a}{2} \dot{\omega}_{\rm pp} t^2} dt\right] \nonumber \\
    =& \frac{2\pi}{s_a m_a} \frac{m_a \omega \omega_{a0}}{\dot{\omega}_{\rm pp, r}} V_{a, r}^2 E_1\simeq E_{\rm mode}^{\rm (inertial)}(\infty),
    \label{eq:g_phi_int}
\end{align}
where we have used Eqs.~(\ref{eq:c_a_dyn}), (\ref{eq:c_a_2_b_a}), and (\ref{eq:g_phi}), and evaluated the integral with a stationary phase approximation on the exponent in Eq. (\ref{eq:c_a_2_b_a}). Note that the resonance condition requires $m_a \omega_r = \omega_a + m_a \Omega$. The tidal torque leads to the post-resonance mode energy, Eq.~(\ref{eq:E_mode_inf}), and it is greater than the absolute value of the interaction energy by a factor of  $\sim (\omega_r t_{\rm gw, r})^{1/2}(r_r/r)^{l+1}$. 
Therefore, the torque term dominates the evolution near mode resonance, and missing the $\gamma_l$ term will lead to an inaccurate description of the dynamical tide. 

\section{Dynamics throughout the inspiral}

Using the tools developed above, we will discuss the coupled evolution of the orbit and the mode in this section. We will start by reviewing the orbital evolution under the PP limit in Sec.~\ref{sec:PP_orb}. Tidal corrections to the orbit will then be derived using both an energy-balancing argument and the osculating orbit technique coupled with the tidal accelerations $g_r^{(t)}$ and $g_\phi^{(t)}$ derived above. In particular, Sec.~\ref{sec:time_phase_shift} is devoted to deriving the phase shifts, the most dominant tidal signature, as functions of both frequency and time. The deviation from a quasi-circular inspiral due to the dynamical tide will be discussed in Sec.~\ref{sec:ecc}. Lastly, we consider additional GW signatures arising from the oscillating f-modes and orbital eccentricity in Sec.~\ref{sec:GW_f_mode}. 

\subsection{Point-particle (PP) evolution}
\label{sec:PP_orb}

We begin the discussion of this section by briefly reviewing the orbital evolution of two PP masses inspiraling due to GW radiation. Results here are used in Sec.~\ref{sec:mode_amp} to derive the linearized mode amplitude evolution. They will also serve as the baseline values onto which we add the tidal perturbation. 

The PP frequency evolution is given by
\begin{equation}
    \dot{\omega}_{\rm pp} = \frac{96}{5}\frac{\eta}{M_t^2} \left(M_t \omega\right)^{11/3}.
    \label{eq:domega_pp}
\end{equation}
We further define $t_{\rm gw} = \omega/\dot{\omega}_{\rm pp}$, which serves as a measure of the instantaneous GW decay timescale. The orbital energy and separation of the PP orbit changes according to 
\begin{align}
    \frac{\dot{E}_{\rm pp}}{E_{\rm pp}} = - \frac{\dot{r}_{\rm pp}}{r_{\rm pp}} = \frac{2}{3}\frac{1}{t_{\rm gw}}, 
\end{align}
where $E_{\rm pp}=-M_1M_2/(2r)$. Note $\dot{E}_{\rm pp}<0$ and $\dot{r}_{\rm pp}<0$ in our convention.

Expressing $\omega$ and $r$ as functions of $t$ leads to
\begin{align}
    &\left(M_t \omega_{\rm pp}\right)^{-8/3} - \left(M_t \omega_0\right)^{-8/3} = \left(\frac{r_{\rm pp}}{M_t}\right)^{4} - \left(\frac{r_0}{M_t}\right)^{4} \nonumber \\
    &=\frac{256 \eta}{5}\left(\frac{t_0}{M_t} - \frac{t}{M_t}\right). 
\end{align}
It is also useful to relate the phase $\phi$ to $\omega$, which has been given earlier in Eq.~(\ref{eq:phi_vs_omega_pp}).

The PP orbital evolution can be modeled by adding the following Burke-Throne accelerations (e.g., Ref. \cite{Flanagan:07}) to the orbital dynamics, Eqs.~(\ref{eq:ddr}) and (\ref{eq:ddphi}),
\begin{align}
    & g_r^{\rm (gw)} = \frac{16 \mu M_t}{5 r^3} \dot{r} \left[\dot{r}^2 + 6 r^2 \dot{\phi}^2 + \frac{4 M_t}{3r}\right], \\
    & g_\phi^{\rm (gw)} = \frac{8 \mu M_t}{5 r^2}\dot{\phi} \left[9 \dot{r}^2 - 6 r^2 \dot{\phi}^2 + \frac{2 M_t}{r}\right]\simeq - \frac{32}{5} \mu r^3 \dot{\phi}^5\label{eq:g_phi_gw}
\end{align}
where we have written the $\phi$ component into a form that is valid even when the $r-\dot{\phi}$ relation is modified by the presence of the tide~\cite{Yu:23a}.






\subsection{Tidal phase shift}
\label{sec:time_phase_shift}

The dominant impact of the tide on the orbital evolution is that it causes a phase shift relative to the PP limit. This section is dedicated to deriving such a phase shift. We will start by deriving first the orbital phase shift as a function of frequency (that is, comparing the phases of two systems at the same frequency). While the phase shift at a given frequency offers a straightforward description of the tide, physically what is measurable is the phase shift measured \emph{at the same time}. Therefore, in the second half of this section, we will also derive the phase shift directly as a function of time using the technique of osculating orbits. The osculating orbits also allow us to extract the post-resonance eccentricity excited by the dynamical tide which we will discuss further in Sec.~\ref{sec:ecc}.  

\subsubsection{Energy balancing}
We use an energy-balancing argument to derive the dynamics as a function of GW frequency $f=\omega/\pi$. The duration of coalescence $t$ and the orbital phase $\phi$ can be computed as 
\begin{align}
    t &= \int \frac{dE/df}{\dot{E}} df, \\
    \phi &= \pi \int f \frac{dE/df}{\dot{E}} df,
\end{align}
Therefore, the tidal corrections to the time and phase can be computed as~\cite{Yu:23a}
\begin{align}
    \Delta t  &\simeq \int \frac{1}{\dot{E}_{\rm pp}}\left( \frac{d\Delta E_{\rm eq}}{d f_{\rm gw}} - \frac{d E_{\rm eq}}{d f_{\rm gw}}\frac{\Delta \dot{E}}{\dot{E}_{\rm pp}} \right) df,
    \label{eq:Delta_t}\\
    \Delta \phi (f) &\simeq \pi \int \frac{f}{\dot{E}_{\rm pp}}\left( \frac{d\Delta E_{\rm eq}}{d f_{\rm gw}} - \frac{d E_{\rm eq}}{d f_{\rm gw}}\frac{\Delta \dot{E}}{\dot{E}_{\rm pp}} \right) df, \label{eq:Delta_phi_vs_f}
\end{align}
where the first term in the parenthesis describes a conservative effect that at a given frequency, the equilibrium energy of the system is modified relative to the PP orbit. We have $\Delta E_{\rm eq} = \Delta E_{\rm pp} + E_{\rm tide}$ and $E_{\rm tide} =  H_{\rm int} + E_{\rm mode}^{\rm (inertial)}$. 
The $\Delta E_{\rm pp}$ describes the change of the orbital energy because both the location and the value of the radial effective potential's minimum change due to the presence of the tide~\cite{Yu:23a},  
\begin{align}
    \frac{\Delta r}{r} \simeq -\frac{g_r^{(\rm t, eq)}}{3 r \omega^2}
    \label{eq:dlogr}, \\
    \frac{\Delta E_{\rm pp}}{E_{\rm pp}} = - 4 \frac{\Delta r}{r}. 
\end{align}
Note that we have used $g_r^{(\rm t, eq)}$, or the equilibrium component of the radial acceleration [using $b_a^{\rm (eq)}$ from Eq. (\ref{eq:b_a_eq}) in Eq. (\ref{eq:g_r})], as it describes the component that is phase coherent with the orbit. When the dynamical tide is excited after mode resonance, the effective potential changes on a timescale shorter than the orbital decay timescale and excites eccentricity of the orbit,  a point we will discuss further in Sec.~\ref{sec:ecc}. The $E_{\rm tide}$ term describes energies due to the tidal degrees of freedom, including both its interaction with the orbit [Eq. (\ref{eq:H_int})] and the inertial-frame mode energy from Eq.~(\ref{eq:E_mode_phy}). 

The tide also affects how much energy is radiated through GW [second term in Eq.~(\ref{eq:Delta_phi_vs_f})]. Here we restrict our discussion to the lowest-order quadrupole radiation that is affected by the tide in two ways. The first one is the $r-\omega$ relation is modified compared to the PP orbit, and the second equality in Eq. (\ref{eq:g_phi_gw}) captures this modification as discussed in Ref.~\cite{Yu:23a}. The second effect is that the tidally induced NS mass quadrupole can couple with the orbital quadrupole, allowing additional GW radiation. 
In our numerical calculations, we do not include the second effect. Instead, we present here an analytical calculation of it. The impact of this term in the adiabatic limit has been well established, see, e.g., Ref. \cite{Yu:23a} and references therein. Thus, we focus on the finite frequency correction (which can still be considered the equilibrium tide in our definition) and the dynamical tide (only excited on resonance) to the NS quadrupole. Our calculation complements the EOB framework which computes the dissipative effect only under the adiabatic (i.e., the zero-frequency) limit; see, e.g., Ref.~\cite{Hinderer:16}. 

From \cite{Yu:23a},
\begin{equation}
    \Delta \dot{E}_{\rm ns-orb} = -\frac{2}{5} \la \dddot{Q}_{\rm ns}^{ij} \dddot{Q}_{\rm orb}^{ij}\ra = -\frac{4}{15} \sum_{m} W_{2m}\la \dddot{Q}^{\rm ns}_{2 m} {\dddot{Q}_{2m}^{\rm orb}}^\ast\ra. 
    \label{eq:dEdt_ns_orb}
\end{equation}
For the orbital mass quadrupole, we have 
\begin{equation}
    \dddot{Q}_{2m}^{\rm orb} \simeq (-im \omega)^3 \mu r^2 e^{-i m \phi};
\end{equation}
the ignored term is smaller by a factor of $1/(\omega t_{\rm gw})$. 

We note that the NS quadrupole can be decomposed into an equilibrium and a dynamical component because
$Q_{2m}^{\rm ns}\propto \left[b_a^{\rm (eq)} + b_a^{\rm (dyn)}\right]$.
The time variation of each component also mainly comes from the phase evolution (respectively goes as $2\omega$ and $2\omega_r$ for the $l=|m|=2$ components), while the rate of change on the amplitude of $b_a$ is smaller by a factor of $\sim \dot{u}/\omega_r\sim (\omega_r t_{\rm gw, r})^{-1/2}$. Furthermore,  
$\dot{b}_a + \dot{b}_a^\ast = 2\Delta_a {\rm Im}[b_a]$, which is small both before resonance (as ${\rm Im}[b_a]$ is small) and during mode resonance as $\Delta_a\simeq0$. 
The equilibrium component is coherent with the orbit throughout the evolution because, in our definition, it is the component that varies at $e^{-i m \phi}$ in the inertial frame. 
The derivation presented in Ref.~\cite{Yu:23a} can thus be readily extended to include the equilibrium tide as
\begin{align}
    \Delta \dot{E}_{\rm ns-orb}^{\rm (eq)} &= - \frac{8}{15}\mu M_1 R_1^2 M_t^{2/3}\omega^{14/3} \nonumber \\
    & \times \sum_{m=\pm2} m^6 W_{2m}\sum_{a, \omega_a>0}^{m_a=m}I_a {\rm Re}[b_a^{\rm (eq)}]. 
\end{align}
Note that 
\begin{equation}
    \frac{\Delta \dot{E}_{\rm ns-orb}(\omega)}{\Delta \dot{E}_{\rm ns-orb}(\omega=0)} = \frac{Q_{22}(\omega)}{Q_{22}(\omega=0)}\neq \kappa_l,
    \label{eq:dE_finite_freq}
\end{equation}
as $m=0$ modes do not contribute to the radiation but they affect the radial acceleration. Therefore, an effective Love number $\kappa_l$ is inaccurate in modeling the radiation from the beat of the NS and orbital mass quadrupoles even in the equilibrium limit. 

We approximate the third derivative of the mass quadrupole due to the dynamical tide as
\begin{equation}
    \dddot{Q}_{22}^{\rm ns, dyn} \simeq (-im_a \omega_r)^3 M R^2 I_a c_a^{\rm (dyn)} e^{-i m_a \phi_r -i m_a \omega_r \tau}. 
\end{equation}
where mode $a$ in the above equation stands specifically for the $l_a=m_a=2$ f-mode with $\omega_a>0$. We also have $\dddot{Q}_{2,-2}^{\rm ns, dyn} =\left(\dddot{Q}_{22}^{\rm ns, dyn}\right)^\ast$.
The energy dissipation rate with the phase expanded near mode resonance reads
\begin{align}
    \Delta \dot{E}_{\rm ns-orb}^{\rm (dyn)} &\simeq -\frac{8}{15} (2\omega_r)^6 \mu r_r^2 M_1 R_1^2 \nonumber \\
    &\times W_{22} I_a 
    {\rm Re}\left[ c_a^{\rm (dyn)} e^{i \dot{\omega}_r \tau^2} \right], 
    \label{eq:dE_dyn}
\end{align}
The phase shift due to this term can be evaluated as 
\begin{align}
    &\Delta \phi_{\rm ns-orb}^{\rm (dyn)} = - \int \omega \frac{\Delta \dot{E}_{\rm ns-orb}^{\rm (dyn)}}{\dot{E}_{\rm pp}} dt,\nonumber \\
    &\simeq \frac{5\pi k_{20}}{192} \left(\frac{M_2}{\eta^2 M_t}\right) \left(\frac{\omega_{a0}}{\omega_r}\right) \left(M_t \omega_r\right)^{5/3} \left(\frac{R_1}{M_t}\right)^5 \nonumber \\
    &\quad \times \left({\rm Fs}\left[\sqrt{\frac{\pi}{2}} u\right] - {\rm Fc}\left[\sqrt{\frac{\pi}{2}} u\right]\right), \nonumber \\
    &\xlongrightarrow{u\to +\infty} 0, 
    \label{eq:dphi_Qns_dyn_Qorb}
\end{align}
where $u=\sqrt{\dot{\omega}_r} \tau$ and we have used a stationary phase approximation in the second line. The mode amplitude is approximated using Eq.~(\ref{eq:c_a_dyn}) at resonance with $F(u=0)\simeq F_{e}(0)$ [see the discussion below Eq.~(\ref{eq:Fe})]. This is justified because $e^{i \dot{\omega}_r \tau^2}$ is even about $\tau=u=0$ while $\left[F_e(u) - F_e(0)\right]$ is odd.
Note that under this approximation, $\Delta \phi_{\rm ns-orb}^{\rm (dyn)}(u\to \infty)=0$. 
While the $F_o(0)$ contribution may survive, it is nonetheless a small correction because $F_o \propto a_3\propto (\omega_r t_{\rm gw, r})^{-1/2}$.  
Therefore, at least at the leading order, the interaction between the NS mass quadrupole due to the dynamical tide and the orbital quadrupole does not lead to net time or phase shift when integrated across mode resonance. 
In the vicinity of mode resonance with $|u|\lesssim 1$ and ${\rm Fs}\neq {\rm Fc}$, the magnitude of tidal phase shift due to this dissipative effect compared to that due to the conservative effect [first term in Eq.~(\ref{eq:Delta_phi_vs_f})] can be estimated from the ratio
\begin{align}
    \frac{\Delta \phi^{\rm (dyn)}_{\rm ns-orb}}{\Delta \phi^{\rm (con)}} &\simeq \Big{|}\frac{\Delta \dot{E}_{\rm ns-orb}^{\rm (dyn)}/\dot{E}_{\rm pp}}{E_{{\rm mode}}^{\rm (inertial)} / E_{\rm orb}} \Big{|}\nonumber \\
    &\sim \frac{\omega_{a0} V_{ar}}{m_a \omega_r |b_a|}\sim \left(\omega_r t_{\rm gw, r}\right)^{-1/2}. 
    \label{eq:dphi_diss_vs_dphi_cons}
\end{align}
The equilibrium component has a similar magnitude because the equilibrium and dynamical mass quadrupoles have comparable magnitude near mode resonance (see the lower panel of Fig. \ref{fig:Q33}). 
Therefore, the conservative effect (energy transfer from the orbit into the excited mode) dominates the phase shift. 



\begin{figure}
    \centering
    \includegraphics[width=0.95\linewidth]{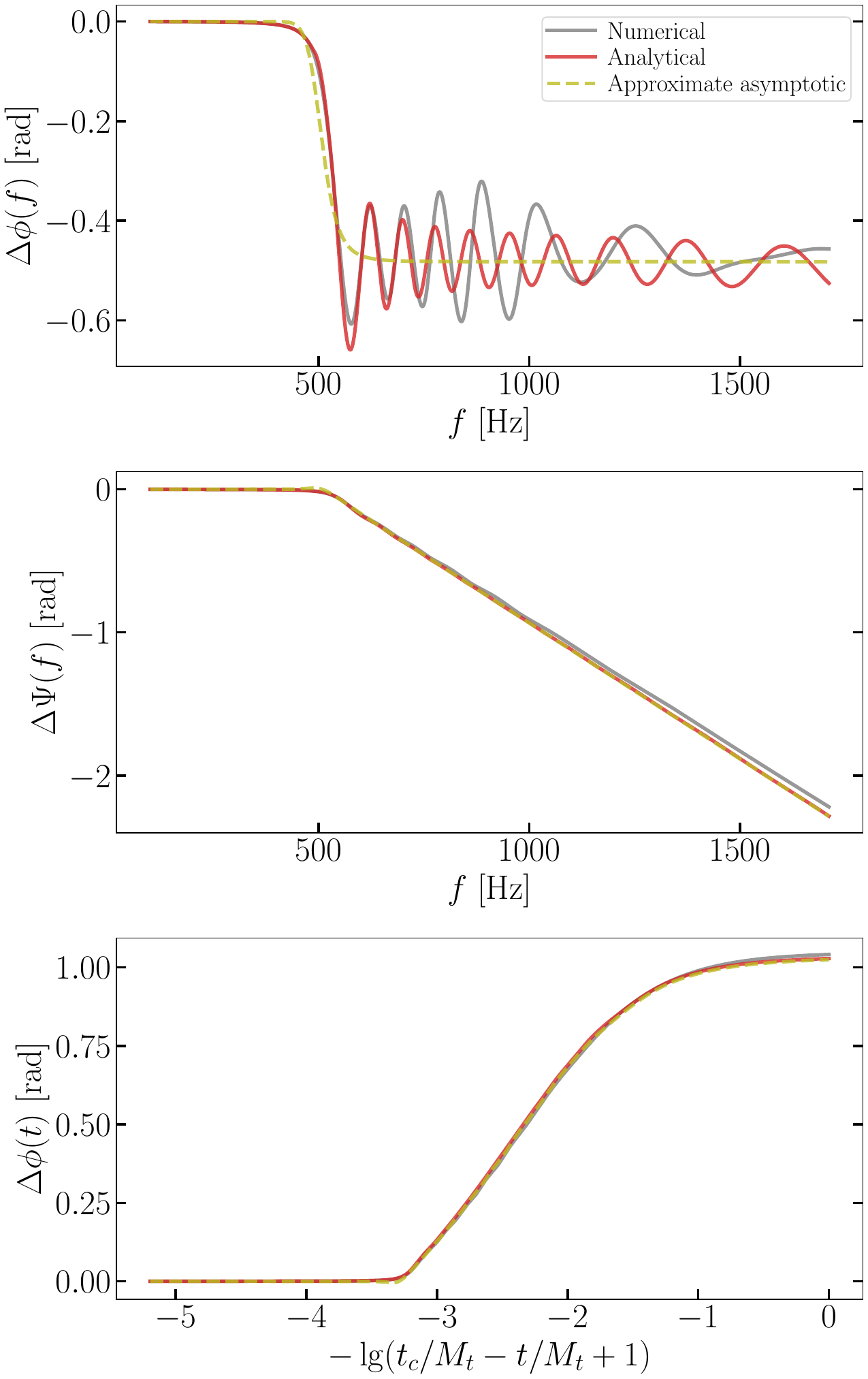}
    \caption{Tidal phase shift due to the $l=3$ f-modes. The top panel shows the orbital phase shift at a given frequency and the bottom one at a given time ($t_c$ is the time when $r=2R$). The middle panel is the GW phase shift of the frequency domain waveform. The gray curves are results extracted numerically while the red curves are from analytical calculations [Eqs. (\ref{eq:Delta_phi_vs_f}), (\ref{eq:Delta_psi}), and (\ref{eq:Delta_phi_vs_t})]. Also shown in the olive-dashed lines are approximate estimations [Eqs.~(\ref{eq:Delta_phi_vs_f_apprx}), (\ref{eq:Delta_psi_vs_f_apprx}), and (\ref{eq:Delta_phi_vs_t_apprx})] that estimate the secular phase shifts.  }
    \label{fig:dphi_l3}
\end{figure}

The phase shift at a given frequency is shown in the top panel of Fig.~\ref{fig:dphi_l3} for the $l=3$ tide. The $l=2$ tide is excluded in the computation here. We use the gray and red lines to respectively represent the numerical and analytical results. It is interesting to note that the main effect can be captured by a step-like jump at mode resonance. 
The amount of jump can be calculated as
\begin{align}
    \Delta t_a = \frac{E_{\rm mode}^{(\rm phy)} (\infty)}{\dot{E}_{{\rm pp}, r}},\text{ and } \Delta \phi_a = \omega_r \Delta t_a,  
    \label{eq:asym_delta_t_phi}
\end{align}
be the asymptotic time and phase shifts of a mode $a$ (and its complex conjugate) integrated to infinity. More explicitly, 
\begin{align}
    \Delta \phi_a &= -\frac{25(2l+1)}{6144 \eta^2} W_{lm}^2 k_{l0} \left(\frac{M_2}{M_1}\right)\left(\frac{R_1}{M_t}\right)^{2l+1} \nonumber \\
    &\times \left(M_t \omega_{a0}\right) \left(M_t \omega_{r}\right)^{(4l-11)/3}. 
\end{align}
The numerical values for the $l_a=3$ and $l_a=2$ f-modes are
\begin{align}
    &\Delta \phi_a\Big{|}_{l_a=3} = -0.48\,{\rm rad}\times \left(4 \eta\right)^{-2} \left(\frac{M_2}{M_1}\right) \left(\frac{k_{30}}{0.02}\right) \nonumber \\
    \times& \left(\frac{R_1}{11.7\,{\rm km}}\right)^{7} \left(\frac{\omega_{a0}/2\pi}{2.5\times10^3\,{\rm Hz}}\right) \left(\frac{\omega_{r}/\pi}{510\,{\rm Hz}}\right)^{1/3}. \\
    &\Delta \phi_a\Big{|}_{l_a=2} = -17\,{\rm rad}\times \left(4 \eta\right)^{-2} \left(\frac{M_2}{M_1}\right) \left(\frac{k_{20}}{0.08}\right) \nonumber \\
    \times& \left(\frac{R_1}{11.7\,{\rm km}}\right)^{5} \left(\frac{\omega_{a0}/2\pi}{1.8\times10^3\,{\rm Hz}}\right) \left(\frac{\omega_{r}/\pi}{950\,{\rm Hz}}\right)^{-1}. 
    \label{eq:dPhi_vs_f_est_l2}
\end{align}

With the jump magnitudes $\Delta t_a$ and $\Delta \phi_a$, we can then write the approximate frequency evolutions of the time and phase shifts as
\begin{align}
& \Delta \bar{t}(f) \simeq \Delta t_a \frac{1 + \tanh u(f)}{2} ,\\
& \Delta \bar{\phi}(f) \simeq \Delta \phi_a \frac{1 + \tanh u(f)}{2},\label{eq:Delta_phi_vs_f_apprx}
\end{align}
where the tanh function is used to smoothly transition the phase shift from nearly 0 to its asymptotic value after mode resonance. The numerical constant in front of $u$ is set to unity for simplicity, yet it provides sufficient accuracy for the approximation, as shown in the olive-dashed line in the top panel of Fig.~\ref{fig:dphi_l3}. 

After a mode's resonance, there appears to be a large fluctuation in the phase shift. Our analytical solution does not fully capture this oscillation because the orbit is no longer quasi-circular and Eq.~(\ref{eq:dlogr}) can capture only the averaged change of separation (see Sec.~\ref{sec:ecc}). Nonetheless, a large portion of the oscillation is not physically observable and therefore does not impact the waveform analysis. For this, we note that phase shift should be measured at a fixed time $t$ instead of frequency $f$ (or $\omega$). They are related through 
\begin{align}
    &\Delta t(f) \simeq - \frac{\Delta \omega(t)}{\dot{\omega}}, \\
    &\Delta \phi(f) \simeq \Delta \phi(t) + \omega \Delta t \simeq \Delta \phi(t) - \frac{\omega}{\dot{\omega}}\Delta \omega(t), \label{eq:dphi_f_vs_dphi_t}
\end{align}
where $\Delta \omega(t)$ and $\Delta \phi(t)$ are the shifts of orbital frequency and of orbital phase measured at a fixed time. 
Since $\omega^2/\dot{\omega}\gg 1$, the second term in the last equality in Eq.~(\ref{eq:dphi_f_vs_dphi_t}) can appear large. It causes a large fluctuation in $\Delta \phi(f)$, though the true observable $\Delta \phi(t)$ varies much less. Equivalently, we can consider the phase of the frequency domain GW waveform, which under the stationary phase is~\cite{Cutler:94}
\begin{equation}
    \Psi(f) = 2 \pi f t(f) - 2\phi(f) - \frac{\pi}{4},
\end{equation}
so the tidal shift is given by\footnote{When comparing the phase shift of two waveforms, it is important to keep in mind that one has the freedom to shift one waveform's time and phase at a reference point relative to the other by arbitrary constants. This is because we do not know a priori of a signal's time and phase of arrival. As a theoretical study, we fix $\Delta t(f_{\rm ref})=0$ and $\Delta \phi(f_{\rm ref})=0$ at $f_{\rm ref}=100\,{\rm Hz}$ for simplicity, which approximately corresponds to $\Delta t(f_{\rm ref}=0)\simeq \Delta \phi(f_{\rm ref}=0)\simeq0$ as the tide is important only when $f\gtrsim 500\,{\rm Hz}$. When computing, e.g., the mismatch between two waveforms~\cite{Lindblom:08}, however, the freedom in $\Delta t(f_{\rm ref})$ and $\Delta \phi(f_{\rm ref})$ needs to be marginalized over.}
\begin{equation}
    \Delta \Psi(f) = 2 \pi f \Delta t(f) - 2\Delta \phi(f)\simeq -2 \Delta \phi(t), \label{eq:Delta_psi}
\end{equation}
and the $\Delta t$ contribution cancels out. 

While $\Delta \phi(f)$ is not directly measurable, it is nonetheless a useful quantity to compute, because its asymptotic behavior can be captured by a particularly simple form, Eq.~(\ref{eq:Delta_phi_vs_f_apprx}). Using Eq. (\ref{eq:dphi_f_vs_dphi_t}) and replacing $\Delta t$ with $\Delta \phi/\omega_r$ [Eq.~(\ref{eq:asym_delta_t_phi})], we have
\begin{align}
    &\Delta \Psi(f) \simeq 2 \left(\frac{\omega}{\omega_r} - 1\right)\frac{1+ \tanh u}{2} \Delta \phi_a, \label{eq:Delta_psi_vs_f_apprx}  \\
    &\Delta \bar{\phi}(t) \simeq \left[1-\frac{\omega(t)}{\omega_r}\right] \frac{1+ \tanh u}{2} \Delta \phi_a, \\
    & =\left[1 - \frac{256\eta}{5} \frac{\tau}{M_t}\left(M_t \omega_r \right)^{8/3}\right]^{-3/8} \frac{1+ \tanh u}{2} \Delta \phi_a, 
    \label{eq:Delta_phi_vs_t_apprx} 
\end{align}
as estimations of the secular shift of the orbital phase in the frequency and time domain, respectively. Note the form is consistent with previous results obtained in, e.g., Refs.~\cite{Flanagan:07, Yu:17b} except for the minor difference that we use here a tanh instead of Heaviside used in Ref.~\cite{Yu:17b} to smoothly connect the pre- and post-resonance values. These approximate expressions are shown in the mid and bottom panels of Fig.~\ref{fig:dphi_l3} in the olive-dashed lines, which show decent agreements with both the numerical (gray lines) and the full analytical (red lines) results. 

\subsubsection{Osculating orbits}
For the completeness of the study, we also present a direct derivation of the dynamics as a function of time $t$. We use the method of osculating orbits and change $(r, \dot{r}, \phi, \dot{\phi})$ to $(p, e, \phi, \phi_0)$, the instantaneous semilatus rectum, eccentricity, orbital phase, and argument of pericenter, according to~\cite{Flanagan:07}
\begin{align}
    &r = \frac{p}{1+ec_\phi}, \label{eq:osc_r} \\
    &\dot{r} = \sqrt{\frac{M_t}{p}}e s_\phi,  \label{eq:osc_dr} \\
    &\dot{\phi} = \sqrt{\frac{M_t}{p^3}}\left[1 + e c_\phi\right]^2, \label{eq:osc_dphi}
\end{align}
where $c_\phi = \cos(\phi - \phi_0)$, $s_\phi=\sin(\phi-\phi_0)$. The inverse relations are 
\begin{align}
    & p = \frac{r^4 \dot{\phi}^2}{M_t}, \\
    & e^2 = \left(\frac{p}{r}-1\right)^2 + \frac{p\dot{r}^2}{M_t}, \\
    & e s_\phi = \dot{r} \sqrt{\frac{p}{M_t}}, \label{eq:osc_esp}\\
    & e c_\phi = \frac{p}{r} - 1. \label{eq:osc_ecp}
\end{align}
The evolution of osculating variables is given by
\begin{align}
    \dot{p} &= \frac{2 \sqrt{p^3}}{\sqrt{M_t} (1+e c_\phi)}g_\phi,  \label{eq:osc_dp}\\
    \dot{e} &= \frac{\sqrt{p}}{2 \sqrt{M_t}(1+e c_\phi)} \left[ (3e + 4c_\phi + e c_{2\phi}) g_\phi \right. \nonumber \\
            &    +\left. (2 s_\phi + e s_{2\phi}) g_r \right], \\
    e\dot{\phi_0} & = \frac{-\sqrt{p}}{2 \sqrt{M_t}(1+e c_\phi)} \left[ (e + 2 c_\phi + e c_{2 \phi}) g_r \right. \nonumber \\
            &    -\left. (4s_\phi + e s_{2\phi}) g_\phi \right] \label{eq:osc_dphi0}
\end{align}
and Eq. (\ref{eq:osc_dphi}).

The zeroth order solution of the set of equations can be obtained from the PP orbit under GW decay 
\begin{align}
    &p_{\rm pp} = r_{\rm pp},\ \left(ec_\phi\right)_{\rm pp}= 0,\ \left(es_\phi\right)_{\rm pp} = -\frac{2}{3}\frac{1}{\omega t_{\rm gw}}.
    \label{eq:osc_0}
\end{align}

To find the tidal correction, we first ignore GW decay and consider a system interacting just due to the equilibrium tide with a radial acceleration $g_r^{\rm (t, eq)}$. Such a conservative system permits a solution with $\dot{p}=\dot{e}=0$ and $c_\phi = 1$, which indicates $e\dot{\phi}_0 = e \dot{\phi}$. From Eqs.~(\ref{eq:osc_dphi}) and (\ref{eq:osc_dphi0}), we have
\begin{equation}
    e c_\phi \simeq - \frac{g_r^{\rm (t, eq)}}{M_t/r_{\rm pp}^2}. 
    \label{eq:ecp_b4_res}
\end{equation}
The same result can also be obtained by first combining Eqs.~(\ref{eq:osc_r}) and (\ref{eq:osc_dphi}) to get 
$
    \dot{\phi}^2 = (M_t/r^3) (1 + e c_\phi),
$
and then compare it with Eq. (\ref{eq:dlogr}) or Eq. (\ref{eq:ddr}) with $\ddot{r}$ set to zero.

Note again the above equation is accurate only before mode resonance; an additional oscillatory component in $ec_\phi$ will be excited by the dynamical tide in the post-resonance regime (Sec.~\ref{sec:ecc}). Nonetheless, Eq.~(\ref{eq:ecp_b4_res}) is sufficient for us to find the deviation in $p$ caused by tide, which follows from Eq.~(\ref{eq:osc_dp}), 
\begin{equation}
    \frac{\Delta \dot{p}}{\dot{r}_{\rm pp}} =  \frac{3}{2} \frac{\Delta p}{r_{\rm pp}} + \frac{\Delta g_\phi}{g_\phi} - e c_\phi.
    \label{eq:d_delp_original}
\end{equation}
The $\Delta g_\phi$ term contains two contributions,
\begin{equation}
    \Delta g_\phi = g_\phi^{(t)} + \Delta g_\phi^{\rm (gw)}. 
\end{equation}
The first piece is due to the tidal torque [mainly from the dynamical tide; Eq.~(\ref{eq:g_phi})]. 
The second piece is because the GW acceleration is modified by the tide. By replacing $(r, \dot{\phi})$ in the second equality in Eq.~(\ref{eq:g_phi_gw}) with osculating variables, we have
\begin{align}
    \frac{\Delta g_\phi^{\rm (gw)}}{g_\phi^{\rm (gw)}} \simeq -\frac{9}{2}\frac{\Delta p}{r_{\rm pp}} + 7 ec_\phi
\end{align}
Consequently, we can rewrite the equation governing the evolution of $\Delta p$ as 
\begin{equation}
    \frac{\Delta \dot{p}}{\dot{r}_{\rm pp}} = \left(-3 \frac{\Delta p}{r_{\rm pp}} + \frac{g_\phi^{\rm (t)}}{g_{\phi}^{\rm (gw)}} - 6 \frac{g_r^{\rm (t, eq)}}{M_t/p_0^2}\right). 
\end{equation}
The solution of this equation is given by 
\begin{align}
    \Delta p(t) &= \frac{1}{r_{\rm pp}^3}\int r_{\rm pp}^3 \dot{r}_{\rm pp} \left(\frac{g_\phi^{(t)}}{g_{\phi0}} - 6 \frac{g_r^{\rm (t, eq)}}{M_t/r_{\rm pp}^2}\right) dt, 
    \nonumber \\
    &=\frac{1}{r_{\rm pp}^3}\int r_{\rm pp}^3 \left(\frac{g_\phi^{(t)}}{g_{\phi0}} - 6 \frac{g_r^{\rm (t, eq)}}{M_t/r_{\rm pp}^2}\right) \frac{dr_{\rm pp}}{d\omega} d\omega,
    \label{eq:dp_vs_t}
\end{align}
Note that while we change the variable to $\omega$ in the second line for easier evaluation of the integral [as the mode amplitudes are given in $u$ and $u$ is treated as a function of $\omega$; Eqs.~(\ref{eq:c_a_dyn}), (\ref{eq:c_a_2_b_a}), (\ref{eq:tau_vs_omega})], here $\Delta p (t)$ is the change in $p$ measured at a fixed time (assuming the tidal and PP orbits are aligned at past infinity). 
After evaluating the integral to a certain frequency $\omega$, we map it back to the desired time according to 
\begin{equation}
    t(\omega)\simeq t_{\rm pp} (\omega) + \Delta t(\omega) \simeq t_{\rm pp} (\omega - \Delta \omega), 
\end{equation}
where $t_{\rm pp}$ is the PP mapping from $\omega$ to $t$, $\Delta t$ is from Eq. (\ref{eq:Delta_t}), and $\Delta \omega$ is computed from Eq. (\ref{eq:Delta_omega}) which we will introduce shortly. 

We can compare the sizes of the two terms in the parenthesis in Eq.~(\ref{eq:dp_vs_t}). First, we note that 
\begin{equation}
    \frac{g_{\phi0}}{M_t/p_0^2} \sim \frac{\dot{r}}{\omega r}.
\end{equation}
Therefore, 
\begin{equation}
    \frac{g_{\phi}^{(t)}/g_{\phi0}}{g_r^{(t)}/(M_t/p_0^2)} \sim \frac{\omega r g_\phi}{\dot{r} g_r},
\end{equation}
which becomes the comparison shown in Eqs.~(\ref{eq:gphi_vs_gr_eq}) and (\ref{eq:gphi_vs_gr_dyn}). In particular, whereas in the equilibrium limit, the radial acceleration plays the major role, near mode resonance it is the torque that dominates the orbit evolution. 

If we ignore the $e c_\phi$ term and adopt the same stationary phase approximation that leads to Eq. (\ref{eq:g_phi_int}), we can approximately carry out the integral in Eq.~(\ref{eq:dp_vs_t}) as
\begin{align}
    \Delta \bar{p}(t) &\simeq -\left(\frac{r_{{\rm pp}, r}}{r_{\rm pp}}\right)^3 \frac{E_{\rm mode}^{\rm (inertial)}(\infty)}{\mu \omega_r^2 r_r} \left[1+\tanh u(t)\right], \nonumber \\
    \simeq& - \dot{r}_{\rm pp}(t)  \Delta t_a \frac{1+\tanh u(t)}{2}. 
    \label{eq:Delta_p_apprx}
\end{align}
The tanh function is introduced to smoothly connect the pre and post-resonance values. 
Alternatively, the same result can be obtained by noticing the change in $p$ at a given frequency $f=\omega/\pi$ is nearly zero as $\Delta p(f) \simeq (4/3) ec_\phi \simeq 0$ based on Eq.~(\ref{eq:osc_dphi}). Eq. (\ref{eq:Delta_p_apprx}) then follows the same argument that leads to Eq.~(\ref{eq:dphi_f_vs_dphi_t}). 

Once we have $\Delta p$ calculated, we can compute the frequency and phase shifts as functions of $t$.
\begin{align}
    &\frac{\Delta \omega}{\omega} = -\frac{3}{2} \frac{\Delta p}{r_{\rm pp}} + 2 ec_\phi \simeq  -\frac{3}{2} \frac{\Delta p}{r_{\rm pp}} - \frac{2g_r^{\rm (t, eq)}}{M_t/r_{\rm pp}^2}, 
    \label{eq:Delta_omega}
    \\ 
    & \Delta \phi(t) = \int \Delta \omega dt \simeq \int \frac{\Delta \omega}{\dot{\omega}_{\rm pp}} d\omega.
    \label{eq:Delta_phi_vs_t}
\end{align}
The resultant $\Delta \phi$ is shown in the red line in the bottom panel of Fig.~\ref{fig:dphi_l3} for the $l=3$ tide. We use $\left[- \lg (t_c/M_t - t/M_t + 1)\right]$ as a modified time coordinate with $t_c$ the time when $r=2R$. Note that when computed as a function of time, the amount of oscillations in the post-resonance part is much milder and it will be quantified in the following section.  

\begin{figure}
    \centering
    \includegraphics[width=0.95\linewidth]{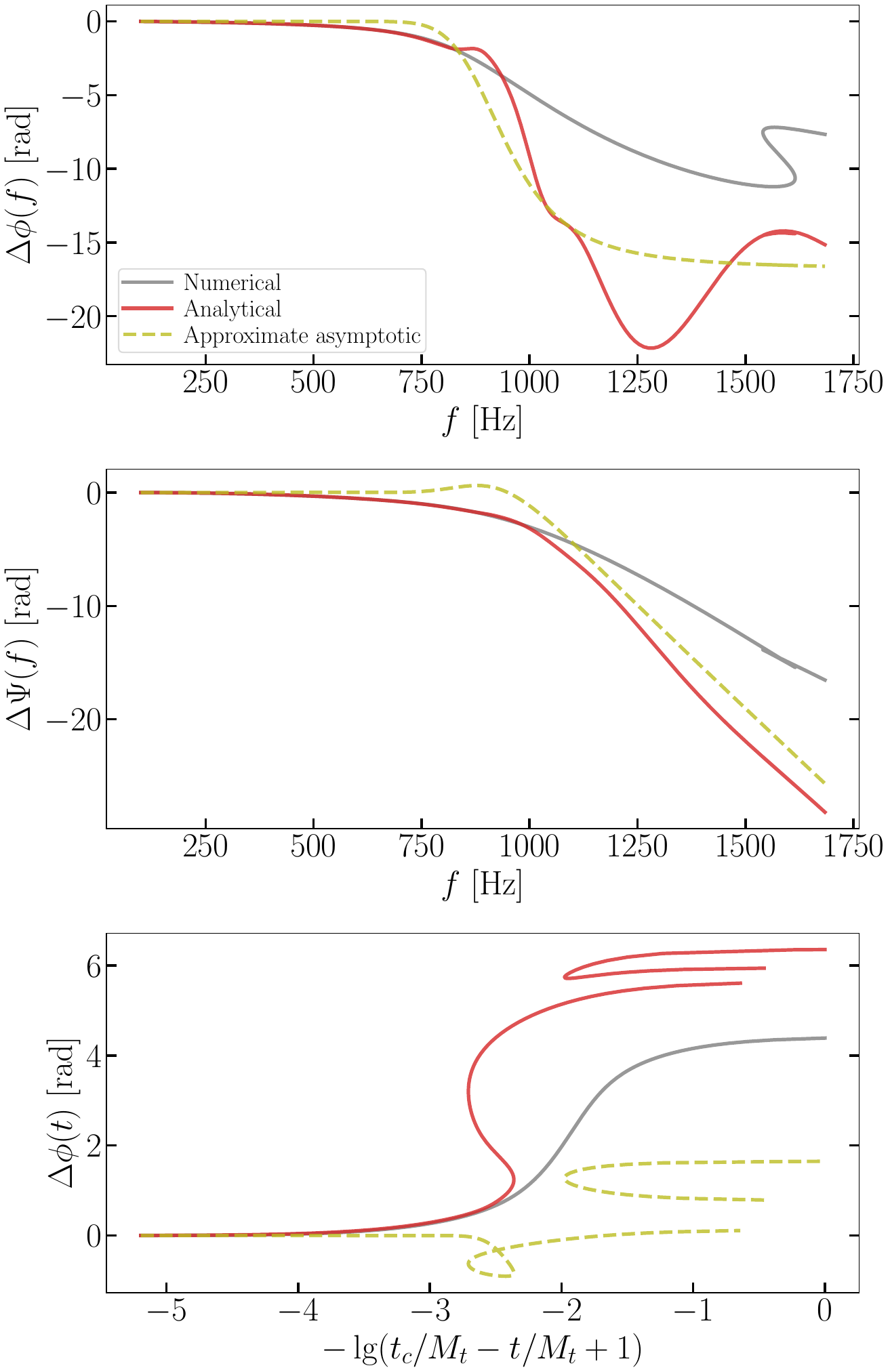}
    \caption{Similar to Fig.~\ref{fig:dphi_l3} but for the $l=2$ tide. The analytical calculations [e.g., Eq.~(\ref{eq:dPhi_vs_f_est_l2})] become inaccurate because the tidal perturbation is so large and keeping only terms linear in $\Delta p/r_{\rm pp}$ is no longer sufficient. }
    \label{fig:dphi_l2}
\end{figure}

As our computation is at the linear order in $\Delta p/p$, it is accurate only when the tide is not too strong. This is why in the figures, we have focused on the $l=3$ tide when comparing the numerical and analytical calculations. The back-reactions are included for the $l=3$ tide and the $l=2$ contributions have been excluded. This does not impact much the calculation of $\Delta \phi_a$ for the $l=3$ f-mode because the $l=3$ is excited at a lower frequency ($\omega_r/(2\pi) = 510\,{\rm Hz}$), or an earlier time, compared to the $l=2$ mode ($\omega_r/(2\pi)=950\,{\rm Hz}$). 
For completeness, we show in Fig.~\ref{fig:dphi_l2} the phase shift induced by the $l=2$ tide (with the tidal back reactions from the $l=3$ components turned off). The analytical estimations [Eq.~(\ref{eq:dPhi_vs_f_est_l2})] start to lose accuracy, emphasizing the need to incorporate higher order effects [which includes both higher order terms in $\Delta p/r$ that come in at $(R/r)^{10}$ and higher order terms in $\xi/ R$ that come in at $(R/r)^{8}$]. 
Interestingly, frequency is no longer a monotonic function of time when the $l=2$ tide is included and this can be seen from the backward turning of the gray line in the top panel of Fig.~\ref{fig:dphi_l2}. This is due to the excitation of orbital eccentricities by the dynamical tide, which we will discuss in Sec.~\ref{sec:ecc} below (see also Fig.~\ref{fig:ecc_l2}). 


\subsection{Eccentricities excited by the dynamical tide}
\label{sec:ecc}


In this section, we discuss the orbital dynamics in the post-resonance part, focusing on the eccentricity excited by the dynamical tide. 

While a non-zero eccentricity $e$ appears in the osculating variables even in the pre-resonance regime, caution is required when interpreting this result. While both the GW decay and the equilibrium tide can cause a finite ``eccentricity'' in the osculating equations (see Eqs. (\ref{eq:osc_0}) and (\ref{eq:ecp_b4_res})), these effects do not break the quasi-circular approximation of the orbit. This is because the argument of pericenter $\phi_0$ proceeds at the same rate as $\phi$ itself. More specifically, we have
\begin{equation}
    \tan (\phi - \phi_0) \simeq \frac{2}{3}\frac{1}{\omega t_{\rm gw}} \frac{M_t/r_{\rm pp}^2}{g_r^{\rm (t, eq)}}, \text{ (equilibrium tide only)}
\end{equation}
which evolves only on the GW decay timescale, much longer than the orbital decay timescale. This means the separation $r$ evolves also slowly over $\sim t_{\rm gw}$. See also Ref.~\cite{Will:19} on related discussions.

\begin{figure}
    \centering
    \includegraphics[width=0.95\linewidth]{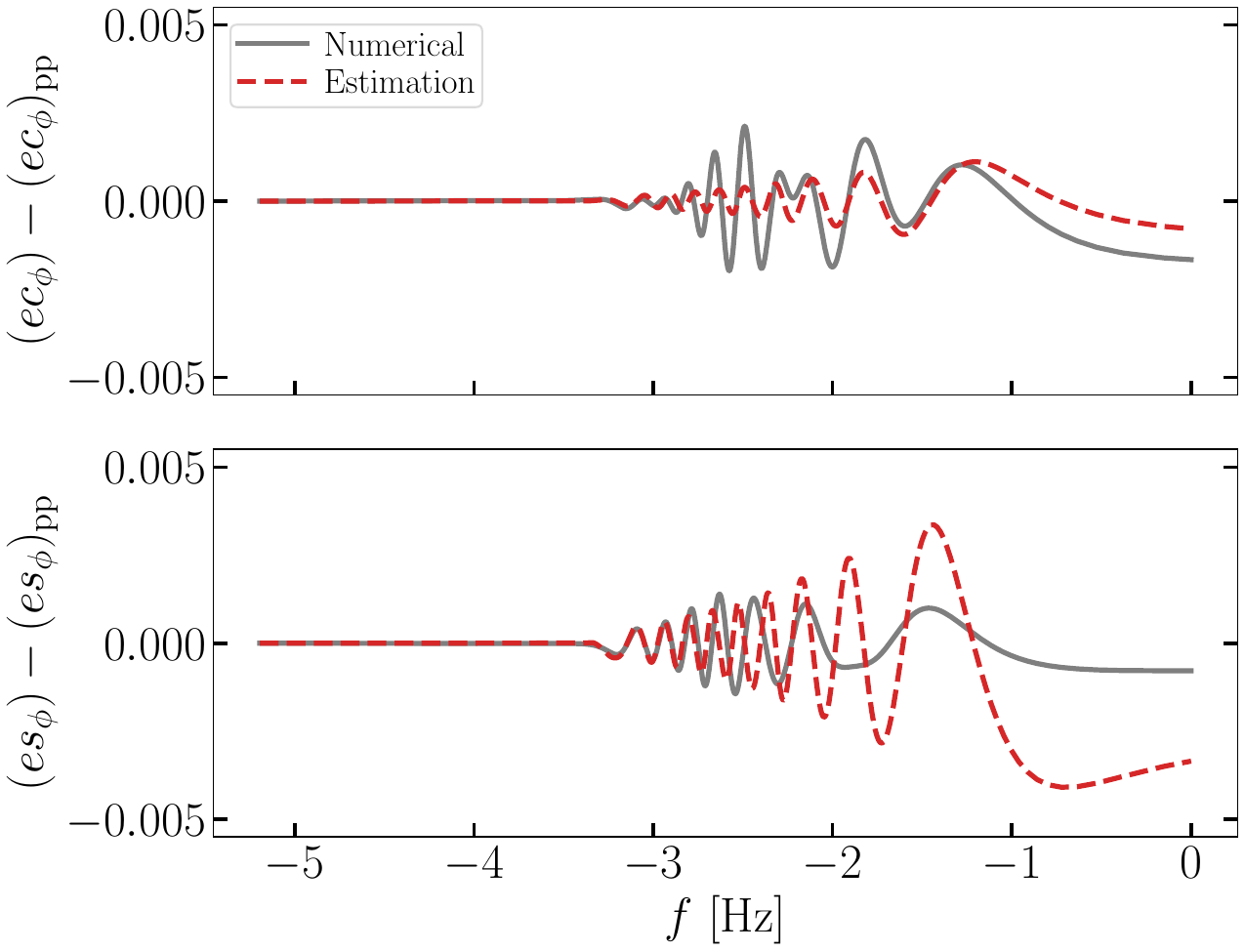}
    \caption{Oscillatory eccentricities in excited by the $l=3$ dynamical tide. }
    \label{fig:ecc_l3}
\end{figure}
\begin{figure}
    \centering
    \includegraphics[width=0.95\linewidth]{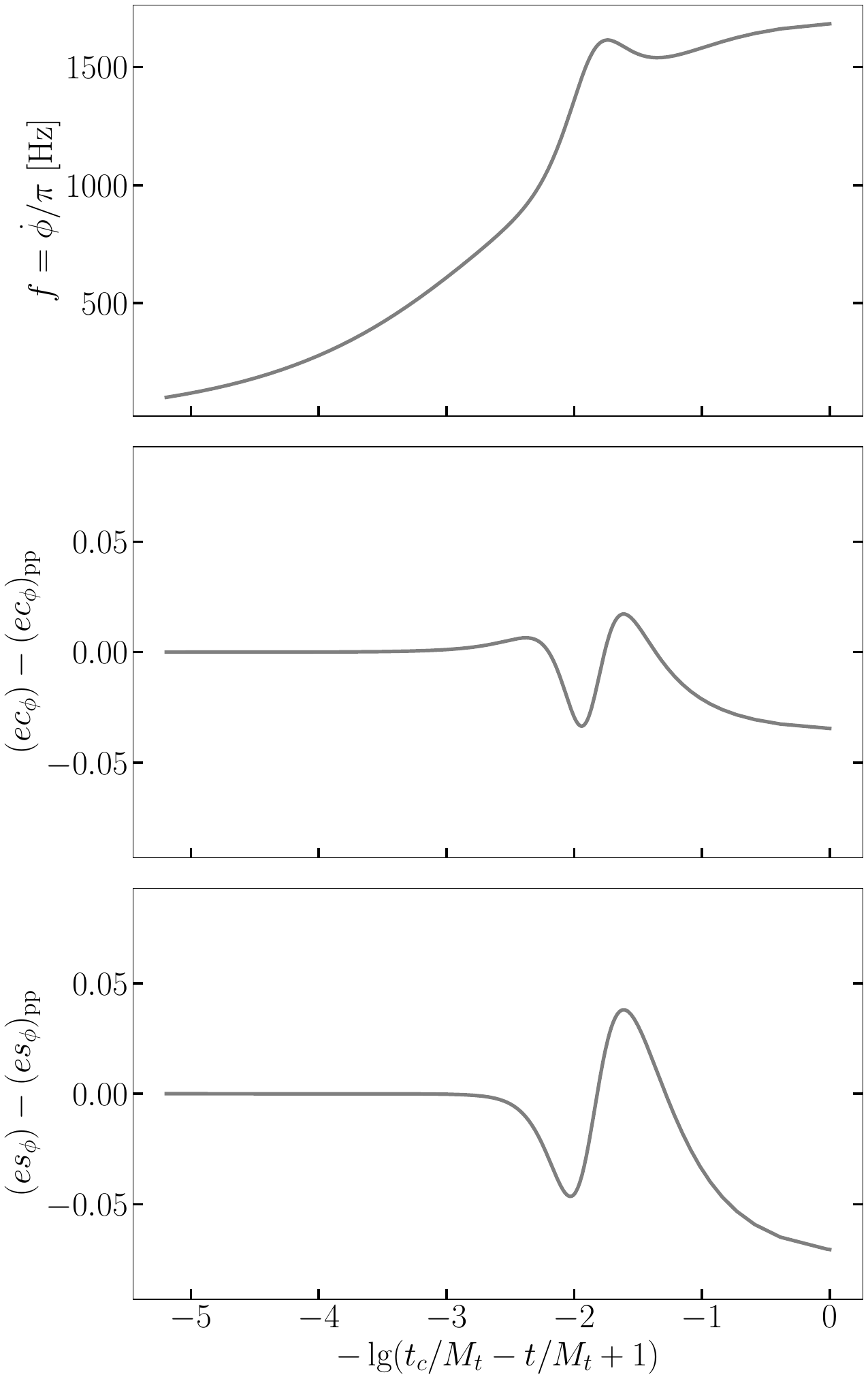}
    \caption{The top panel shows the frequency-time relation when the $l=2$ tide is included. The frequency is computed $f=\dot{\phi}/\pi=\omega/\pi$. The middle and bottom panels are similar to Fig.~\ref{fig:ecc_l3} and it shows the eccentricity excited by the $l=2$ dynamical tide. All results shown in this figure are extracted numerically.  }
    \label{fig:ecc_l2}
\end{figure}

In contrast, once a mode is excited, the interaction energy changes on a timescale $[m(\omega - \omega_r)]^{-1}$, which can be much faster than the GW decay timescale. Therefore, the orbit cannot settle at the bottom of the effective potential and therefore cannot remain quasi-circular due to the interaction energy being oscillatory.  Instead, a \emph{forced} eccentricity is excited~\cite{Murray:00}. 
The results are numerically confirmed in Figs.~\ref{fig:ecc_l3} and \ref{fig:ecc_l2}.
To quantify this, we note that the binary separation $r$ and its time derivative $\dot{r}$ do not depend on $e$ alone but instead depend on $e c_\phi$ and $e s_\phi$. Our goal is therefore to find approximate expressions for the oscillatory parts in $e c_\phi$ and $e s_\phi$ as they are the terms causing deviations to the quasi-circular approximation.

To proceed, we start with the assumption that $e c_\phi \simeq \Delta p/r_{\rm pp}$ is small. This assumption will be justified later. To estimate $ e s_\phi$ then, we note from Eq.~(\ref{eq:osc_dr}) that $e s_\phi \propto \dot{r} \simeq \dot{p}$, and from Eq.~(\ref{eq:osc_dp}) we have $\dot{p} \propto g_{\phi}$. Therefore, we arrive at 
\begin{equation}
    (e s_\phi)^{\rm (osc)} \simeq -\frac{g_\phi^{(t)}}{g_\phi^{\rm (gw)}} (e s_\phi)_{\rm pp} =  2 \frac{g_\phi^{(t)}}{M_t/r_{\rm pp}^2}, \label{eq:esp_dyn_tide}
\end{equation}
where we have used the superscript ``(osc)'' to emphasize that it is the oscillatory component, in contrast to the secular component in Eq.~(\ref{eq:osc_0}). 
To estimate $(e c_\phi)^{\rm (osc)}$, we note that $(\phi - \phi_0)\simeq -3\pi/2$, so 
\begin{align}
    &\frac{d}{dt}(e c_\phi)^{\rm (osc)} \simeq e_{\rm pp} (\omega - \dot{\phi}_0), \nonumber \\
    \simeq& \omega \left[- (e s_\phi)_{\rm pp}  + 2  \frac{g_\phi}{M_t/r_{\rm pp}^2}\right] \simeq \frac{2}{\omega r_{\rm pp}} g_\phi^{(t)}.
\end{align}
Integrating the equation and approximate quantities varying on the GW decay timescale as constants [or equivalently, doing integration by parts and ignoring terms suppressed by $1/(\omega t_{\rm gw})$], we have
\begin{align}
    &(e c_\phi)^{\rm (osc)} \simeq \frac{2}{\omega r_{\rm pp}}\int g_\phi^{(t)} dt,
\end{align}
Further, 
\begin{align}
    &g_\phi^{(t)} \simeq - \frac{1}{(l_a+1)} \frac{\dot{g}_r^{(t)}}{\omega-\omega_r},
    \label{eq:g_phi_vs_dg_r_dt}, \\
    &g_r^{(t)} \simeq  \frac{l_a+1}{m_a^2} \frac{\dot{g}_\phi^{(t)}}{\omega-\omega_r}.
    \label{eq:g_r_vs_dg_phi_dt},
\end{align}
in the post-resonance regime. 
Using Eq. (\ref{eq:g_phi_vs_dg_r_dt}) to evaluate $\int g_\phi^{(t)} dt$, we have
\begin{equation}
    (ec_\phi)^{\rm (osc)} \simeq -\frac{2}{l+1} \frac{g_r^{(t)}}{M_t/r_{\rm pp}^2},
    \label{eq:ecp_dyn_tide}
\end{equation}
where we have replaced the $(\omega-\omega_r)$ part in Eq.~(\ref{eq:g_phi_vs_dg_r_dt}) by $\omega$ to avoid divergence near resonance and it does not affect much the post-resonant estimations. Note that $(ec_\phi)^{\rm (t,osc)} / (es_\phi)^{\rm (t,osc)} \simeq l/(l+1)^2 < 1$, which is consistent with the assumptions we made at the beginning. 

The comparison between the numerical and analytical estimations of the oscillatory eccentricity is shown in Fig.~\ref{fig:ecc_l3} for the $l=3$ tide (with $l=2$ tide excluded). To numerically extract the oscillatory component, we use Eqs.~(\ref{eq:osc_esp}) and (\ref{eq:osc_ecp}) to compute the total eccentricities from the numerical $(r, \dot{r}, \phi, \dot{\phi})$ and then remove the PP values given in Eq.~(\ref{eq:osc_0}). The equilibrium tide can also cause a secular contribution to $e c_\phi$ [Eq.~(\ref{eq:ecp_b4_res})], yet numerically it is sufficiently small to be ignored. The analytical estimations are accurate for a small range of frequencies after mode resonance and lose accuracy in the later evolution. Nonetheless, they provide a decent estimation of the order of magnitude for the deviation from the quasi-circular approximation. Interestingly, the magnitude of the eccentricities increases with frequency as the magnitude of $g_{r, (\phi)}^{(t)}$ increases. 

The numerically extracted eccentricity due to the $l=2$ tide is shown in Fig.~\ref{fig:ecc_l2}. Here in the top panel, we present also the instantaneous GW frequency (still defined through $f=\dot{\phi}/\pi$) as a function of time, which is no longer a monotonically increasing function as in the PP case. This confirms the eccentricity excitation as shown in the middle and bottom panels. The magnitude of this eccentricity can exceed 0.05 near the end of the inspiral. An estimation of its detectability will be presented later in Sec.~\ref{sec:GW_f_mode}.

Lastly, we estimate the magnitude of the oscillatory eccentricity as a function of the phase shift $\Delta \phi_a$ [Eq.~(\ref{eq:asym_delta_t_phi})]. We have
\begin{align}
    &\Delta E_{\rm eq} \simeq \omega_r \mu r_r g_\phi^{(t)} t_{\rm res}, \\
    &\Delta \phi \simeq \omega_r \frac{\Delta E_{\rm eq}}{\dot{E}_{\rm pp}}, 
\end{align}
where $t_{\rm res}= \sqrt{1/\dot{\omega}_r}$ is the duration of resonance. 
We thus have
\begin{equation}
    e s_\phi  \sim |\Delta \phi| \left(\frac{r_r}{r}\right)^{l} (\omega_r t_{\rm gw, r})^{-3/2}. 
    \label{eq:ecc_vs_dphi}
\end{equation}
While the magnitude of the forced eccentricity increases as $1/r^{l}\propto \omega^{2l/3}$, its overall magnitude is suppressed by the large factor of $(\omega_r t_{\rm gw, r})^{3/2}\propto \omega_r^{-5/2}$. The post-resonance eccentricity is therefore less significant compared to the phase shift for modes excited earlier during the inspiral (e.g., gravity and inertial modes). 

\subsection{GW from the dynamical tide}
\label{sec:GW_f_mode}

While the dominant impact of the tide is the phase shift discussed in Figs.~\ref{fig:dphi_l3} and \ref{fig:dphi_l2}, we also consider additional features caused by the dynamical tide. In particular, once the f-mode is resonantly excited, it will ring at its natural frequency ($=m_a \omega_r$ in the inertial frame) and emit GW accordingly. 
This is similar to the post-merger oscillation of two non-spinning NSs~\cite{Read:09, Hotokezaka:13, Takami:14, Takami:15, Rezzolla:16,Kawamura:16, Dietrich:18b, Zappa:18, Most:22}. For the oscillation to be excited during the inspiral stage, it requires either a highly eccentric orbit or a rapid, anti-aligned spin of the NS. Numerical simulations \cite{Huan:18, Vivekanandji:18} have confirmed the former case, and our study aims to provide a detailed discussion of the latter scenario when the NS is rapidly spinning. 
Further, the deviation from the quasi-circular approximation (Fig.~\ref{fig:ecc_l2}) allows GW to be emitted at frequencies other than $\dot{\phi}/\pi$ for the quadrupole GW. Both effects are examined in this section for the $l=2$ tide that dominates the back reaction. 

To compute the GW strain from the oscillating f-mode, we use the quadrupole formula~\cite{Poisson:14},
\begin{equation}
    h^{ij} = \frac{2}{D_L} \ddot{Q}^{\la ij \ra}. 
\end{equation}
From Eq. (\ref{eq:multipole_from_amp}), we have
\begin{align}
    \ddot{Q}^{ij}_{\rm ns} &\simeq -2 N_2 (2 \omega_r)^2 M_1 R_1^2 I_a \nonumber \\
    &\times {\rm Re}\left[\left(\mathcal{Y}_{22}^{ij}\right)^\ast c_a e^{-2i\omega_r t + i\psi_r}\right], 
\end{align}
where we have expressed the mode amplitude in terms of $c_a = q_a \exp[ i (\omega_a t - \psi_r)]$ which stays nearly constant after resonance. The subscript $a$ stands for the f mode with $(l_a, |m_a|)=(2, 2)$ and the factor of $2$ comes from summing over $m_a=\pm 2$ contributions.

The strain tensor can be projected to the two polarizations as
\begin{align}
    &h_+ = \frac{1}{2}(e_X^i e_X^j - e_Y^i e_Y^j)h_{ij}, \\
    &h_\times = \frac{1}{2}(e_X^i e_Y^j + e_Y^i e_X^j)h_{ij}, 
\end{align}
where $\vect{e}_X {=} \left(\cos \psi, -\sin \psi, 0 \right)$, $\vect{e}_Y {=} \left(\cos \iota \sin \psi, \cos \iota \cos\psi, -\sin\iota \right)$. The angle $\iota$ is the line of sight and $\psi$ specifies the x-axis of the detector relative to the orbit. 
We have 
\begin{align}
    &h_+^{\rm ns} = -\frac{(1 + \cos^2 \iota)}{2} \left(\frac{2\omega_r}{\omega_1}\right)^2 h_0^{\rm ns}
    {\rm Re}\left[ c_a e^{-i2\omega_r t + i\psi_0}\right], \label{eq:hp_ns}\\
    &h_\times^{\rm ns} = \cos \iota  \left(\frac{2\omega_r}{\omega_1}\right)^2 h_0^{\rm ns}
    {\rm Im}\left[ c_a e^{-i2\omega_r t + i\psi_0}\right],\label{eq:hc_ns}
\end{align}
where $\omega_1^2 = M_1/R_1^3$, $\psi_0 = \psi_r - 2 \psi$, and
\begin{equation}
    h_0^{\rm ns} = \sqrt{\frac{32 \pi}{15}} \frac{ M_1^2}{R_1 D_L}  I_a. 
\end{equation}
For the rest of the discussions, we will assume that the detector's antenna response is such that the observed strain is given by $h_+$. 
To gain some analytical insights, we can replace $c_a$ with the asymptotic mode energy, Eq.~(\ref{eq:E_mode_inf}), leading to 
\begin{equation}
    h_+^{\rm ns} \simeq \frac{(1 + \cos^2 \iota)}{2} A_{\rm ns} \cos (2\omega_r t + \psi_0'), 
\end{equation}
where 
\begin{align}
    A_{\rm ns} &\simeq \sqrt{\frac{\omega_r t_{\rm gw, r}}{4\pi}} \left(\frac{M_2}{M_t}\right) \left(\frac{\omega_{a0}}{\omega_1}\right)\left(\frac{2\omega_r}{\omega_1}\right)^3  \frac{k_{20} M_1^2}{R_1 D_L}\nonumber \\
    &\propto \omega_r^{13/6}. 
\end{align}
While for the $l=2$ tide, Eq.~(\ref{eq:E_mode_inf}) loses accuracy because of the neglecting of higher-order effects, it nonetheless provides the scaling of the strain amplitude with different parameters, in particular the frequency, or $\omega_r$, at which the mode is resonantly excited, hence the spin rate $\Omega_s$ as $m_a \omega_r = \omega_a + m_a \Omega_s$. 
The Fourier transform is given by (with $f>0$)
\begin{align}
    &|\tilde{h}_+^{\rm ns}(f)| = \Big{|}\int h_+^{\rm ns}  e^{i 2\pi ft} dt\Big{|} \nonumber \\
    &\simeq \Big{|} \frac{(1 + \cos^2 \iota)}{2} \frac{A_{\rm ns}T}{2} {\rm sinc}\left[(2\pi f - 2 \omega_r)\frac{T}{2}\right] \Big{|},
    \label{eq:hp_ns_f_domain}
\end{align}
where $T\simeq (3/8)t_{\rm gw, r}$ is the approximate duration for the f-mode to ring, so $|\tilde{h}_+^{\rm ns}(f)|\propto \omega_r^{-1/2}$. 
This allows us to compute the signal-to-noise ratio (SNR $\rho$) of the f-mode, 
\begin{equation}
    \rho^2 = \int \frac{4 f \tilde{h}^\ast(f) \tilde{h}(f) }{S(f)} d\log f,
\end{equation}
which varies slowly with respect to $\omega_r$ when the noise power spectral density (PSD) $S(f)$ is nearly flat. 

Damping on the excited f-mode has been ignored in the calculations above. This is justified by noticing that the energy loss from the excited f-mode can be estimated as [assuming it is dominated by GW radiation and other mechanisms, such as Urca reactions~\cite{Arras:19} or nonlinear fluid instabilities~\cite{Weinberg:16}, are subdominant; 
cf. Eq. (\ref{eq:dEdt_ns_orb})]
\begin{equation}
    \dot{E}_{\rm mode} = -\frac{8}{15} W_{22} \la \dddot{Q}^{\rm ns}_{22} {\dddot{Q}_{22}^{\rm ns}}^\ast\ra. 
\end{equation}
The energy-damping timescale of the excited f-mode is thus
\begin{align}
    &t_{\rm mode} = \frac{E_{\rm mode}^{\rm (inertial)}(\infty)}{|\dot{E}_{\rm mode}|} \nonumber \\
    &=\frac{3\pi}{64}\frac{1}{W_{22}k_{20}} \left(\frac{\omega_1}{\omega_r}\right)^{5} \left(\frac{\omega_1}{\omega_{a0}}\right)\left(M_1 \omega_1\right)^{-5/3} \omega_1^{-1}. 
\end{align}
The ratio of this timescale to the remaining time of the inspiral is thus
\begin{align}
    \frac{t_{\rm mode}}{\frac{3}{8}t_{\rm gw, r}} = \frac{12\pi \eta}{5 W_{22}k_{20}} \left(\frac{M_t}{M_1}\right)^{5/3} \left(\frac{\omega_1}{\omega_r}\right)^{7/3} \left(\frac{\omega_1}{\omega_{a0}}\right). 
\end{align}
Numerically, this ratio is about $1.4\times10^3$ for the system we consider. Therefore, damping of the f-mode can be safely ignored. 

\begin{figure}
    \centering
    \includegraphics[width=0.95\linewidth]{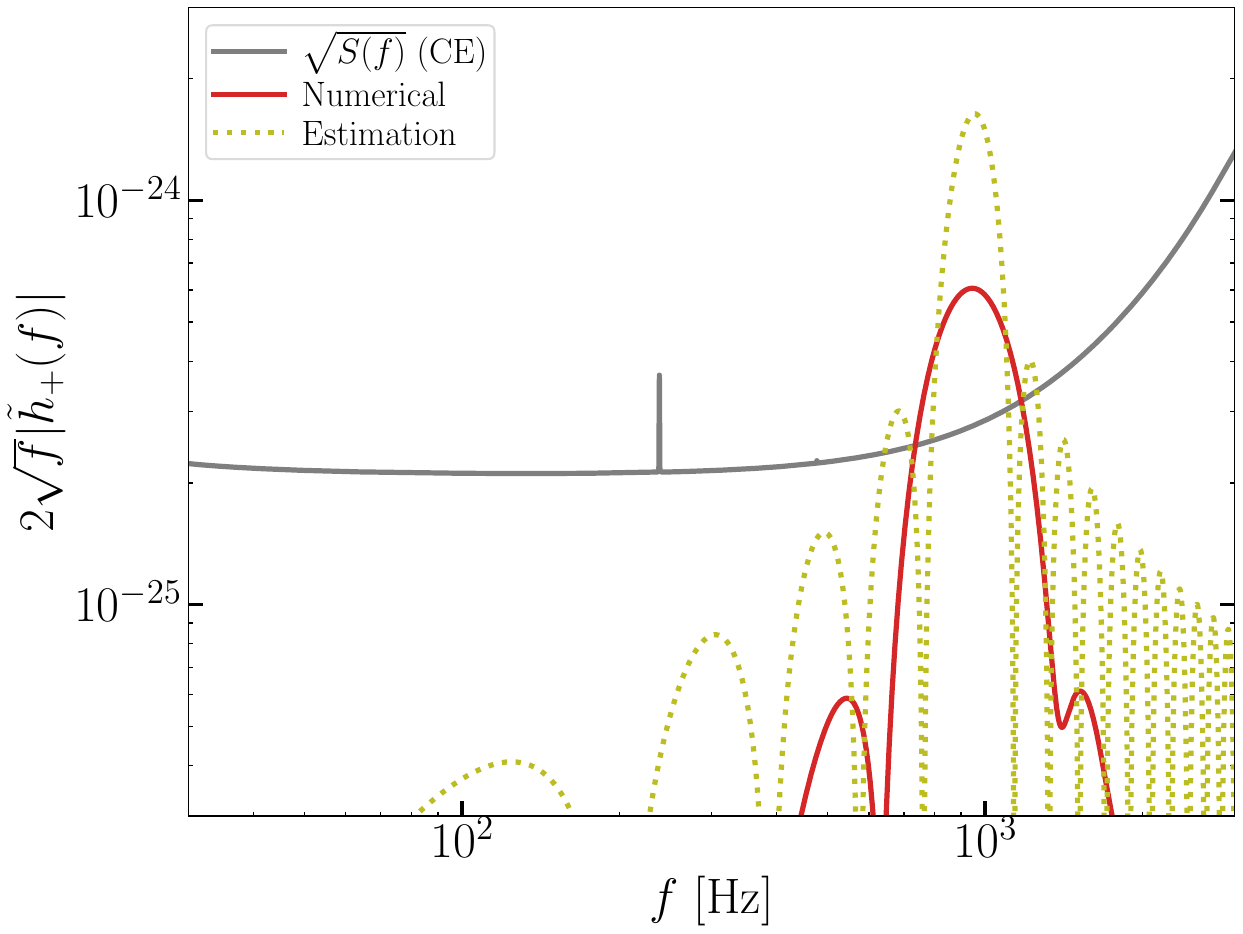}
    \caption{GW strain from the $l_a=|m_a|=2$ NS f-mode excited by the dynamical tide, assuming the source is located at $D_L=50\,{\rm Mpc}$ and the detector's antenna response is such that the observed strain is $h_+$. The red line is computed using the numerical mode amplitude and the olive-dotted line is from the analytical estimation Eq.~(\ref{eq:hp_ns_f_domain}). The sensitivity of the next generation of GW detector Cosmic Explore is shown in the gray line. The numerically computed strain can be detected with a matched-filtering SNR of $\rho=1.3$}. Varying the spin rate of the NS does not affect the SNR much.
    \label{fig:h_ns}
\end{figure}

In Fig.~\ref{fig:h_ns} we show the strain from the $l_a=|m_a|=2$ NS f-mode excited by the dynamical tide. The red trace is computed by plugging into Eq.~(\ref{eq:hp_ns}) the numerical mode amplitude $c_a$ with the equilibrium component [Eq.~(\ref{eq:b_a_eq})] removed. The source is placed at a luminosity distance of $D_L=50\,{\rm Mpc}$. The time domain waveform is sampled at a rate of 8192 Hz and then transferred to the frequency domain. To avoid boundary effects in the Fourier transform, we let the system evolve to $r=R_1$ and then window out the portion where $r<2.3 R_1$ with half of a Hann window (i.e., the window function varies from 1 at time when $r=2.3 R_1$ to $0$ when $r=R_1$). 
The GW frequency corresponding to the truncation is $1390\, {\rm Hz}\simeq 0.89 f_{\rm isco}$. 
The SNR estimate is insensitive to the truncation location because the amplitude of the signal in the frequency domain is proportional to the duration of the signal, and $T\sim t_{\rm gw}\sim f^{-8/3}$ is dominated by the low-frequency end (i.e., the location of resonance). 
If we instead more conservatively window out the portion where $r< 2.6 R_1$, or $f > 1160\,{\rm Hz}$, the SNR is reduced only by 7\%. 
As a reference, the olive-dotted line shows the analytical estimation based on Eq. (\ref{eq:hp_ns_f_domain}), which overestimates (underestimates) the amplitude (width) the width of the peak. We nonetheless verified its accuracy by artificially reducing the Love number by a factor of 100 to eliminate higher-order $(\Delta r/r)$ effects.  Also plotted in the gray line is the sensitivity of Cosmic Explorer (CE)~\cite{Evans:17, Evans:19, Evans:21}. 

Interestingly, the GW from the f-mode can be detected with an SNR $\rho=1.3$ for a source at 50\,Mpc from the numerical result (red curve in Fig.~\ref{fig:h_ns}). Because CE's sensitivity is nearly flat up to 1000 Hz, the SNR does not vary much concerning the spin rate of the NS as long as it is more negative than the critical value estimated in Eq.~(\ref{eq:Omega_crit}), consistent with the scaling discussed below Eq.~(\ref{eq:hp_ns_f_domain}). At $\Omega=-2\pi \times 300\,{\rm Hz}$, we can still recover the signal with $\rho=0.8$. 
If the NS is described by the harder H4 EOS, then the SNR becomes $\rho = 0.9$ when $\Omega=-2\pi \times 200\,{\rm Hz}$, and $\rho=2$ when $-\Omega\gtrsim 2\pi \times 400\,{\rm Hz}$. 
This indicates GW from the NS f-mode can be a signature to look for in addition to GW from the orbit with the next generation of detectors (including CE and the Einstein Telescope~\cite{Hild:10, Sathyaprakash:12}, as well as the Neutron Star Extreme Matter Observatory~\cite{Ackley:20} that has a sensitivity comparable to CE with $f{\sim}\,\text{a few} \times 1000\,{\rm Hz}$). The peak-like feature of this signal should allow it to be detected with little degeneracy with other effects from the orbital inspiral, and identifying the peak frequency directly constrains the combination of NS f-mode frequency and its spin rate, while the height provides additional information on the Love number. 

The deviation of a quasi-circular inspiral is another effect caused by the dynamical tide that we want to examine. For this, we decompose the strain tensor into modes as
\begin{equation}
    h_+ - i h_\times = \sum_{m}h_{2m} {}_{-2}Y_{2m},
\end{equation}
where ${}_{-2}Y_{2m}$ is spin weighted spherical harmonic with spin $-2$. When computing the derivatives of the orbital quadrupole $Q_{ij}^{\rm orb} = \mu r_i r_j$, we replace $\ddot{r}$ and $\ddot{\phi}$ using Eqs.~(\ref{eq:ddr}) and (\ref{eq:ddphi}), leading to
\begin{align}
    h_{22} &= 4 \sqrt{\frac{\pi}{5}} e^{-2i\phi} \frac{\eta M_t}{D_L}\left[\frac{M_t}{r} + r^2 \dot{\phi}^2 \right.  \nonumber \\
    &\left.- r g_r -\dot{r}^2 + i r\left(g_\phi + 2\dot{r}\dot{\phi}\right)\right]. 
\end{align}
In terms of the osculating variables, we have
\begin{align}
    &h_{22} = h_0 e^{-2i\phi}\left[1  +  \left(\frac{3}{2} e c_\phi + i e s_\phi\right)  + \frac{(-g_r + i g_\phi)}{2M_t/ p^2}\right], 
\end{align}
where 
\begin{align}
    &h_0 = 8 \sqrt{\frac{\pi}{5}}\frac{\mu M_t}{D_L p}. 
\end{align}
Because $g_{r(\phi)}^{(t)}/(2M_t/ p^2) \sim (e s_\phi)^{\rm (osc)}$ [Eq.~(\ref{eq:esp_dyn_tide})], the non-circular component of the GW can be estimated as
\begin{equation}
    \delta h_{22} \sim (e s_\phi)^{\rm (osc)} h_0 e^{-2i\phi}.
\end{equation}
Because $(e s_\phi)^{\rm (osc)} \sim \cos \left[m_a (\phi -  \omega_r t)\right]$, we see that $\delta h_{22}$ has a component that varies at $2\omega_r$ (together with a high-frequency component at $4\omega - 2 \omega_r$ which is outside the sensitivity band of a GW detector). Therefore, this effect should be added coherently with the GW from the NS f-mode [Eq. (\ref{eq:hc_ns}) and (\ref{eq:hp_ns})].  

It is thus interesting to compare $D_L \delta h_{22}$ and $h_{22}^{\rm (ns)}$. 
The GW strain from the oscillatory orbital eccentricity is
\begin{align}
    D_L \delta h_{22} \sim D_L (e s_\phi)^{\rm (osc)} h_0 \sim E_1 V_a b_a,
\end{align}
where we have replaced $(e s_\phi)^{\rm (osc)}$ in terms of $g_\phi^t$ using Eq.~(\ref{eq:esp_dyn_tide}) and then used Eq. ({\ref{eq:g_phi}}). 
On the other hand, the GW strain from excited NS mode is 
\begin{align}
    D_L h_{22}^{(\rm ns)} \sim \omega_a^2 b_a I_a M_1 R_1^2 \simeq E_1 I_a b_a. 
\end{align}
Therefore, $D_L \delta h_{22} \sim (R_1/r)^3 D_L h_{22}^{(\rm ns)}$, and the GW from the orbital eccentricity is subdominant compared to the GW from NS f-mode itself.



\section{conclusion and discussion}
\label{sec:conclusion}

The key conclusion of the paper is the following.

(1). We showed a new approach to analytically solve the time evolution of mode amplitude, or equivalently NS mass multipoles [Eq.~(\ref{eq:multipole_from_amp})], in the presence of resonance, by decomposing the amplitude into a resummed equilibrium component [including finite frequency corrections; Eq.~(\ref{eq:b_a_eq})] and a dynamical component that is excited only around mode resonance [Eq.~(\ref{eq:c_a_dyn})]. The new formalism simplifies the numerical implementation as both components remain finite throughout the evolution. This avoids the need to subtract two diverging terms at mode resonance as required in previous analysis \cite{Steinhoff:16, Hinderer:16, Steinhoff:21}. Furthermore, the solution can be extended in the post-resonance regime whereas previous works are accurate only up to the resonant frequency (Fig.~\ref{fig:Q33}). 

(2). The effective Love number proposed in, e.g., Refs.~\cite{Hinderer:16, Steinhoff:21} ($\kappa_l$ in our notation) can capture the radial backreaction [Eq.~(\ref{eq:g_r})] but it does not account for the tidal torque [Eq. (\ref{eq:g_phi})]. To capture the torque, an additional dressing factor $\gamma_l$ originating from the imaginary part of the Love number of each $(l, m)$ harmonic is necessary [Eq. (\ref{eq:gamma_l})]. Near mode resonance, it is the tidal torque that dominates the impact on the orbit (Fig.~\ref{fig:Etide}). The effective Love number is also insufficient to describe the finite-frequency correction to the GW radiation from the interaction of NS and orbital mass multipoles [Eq.~(\ref{eq:dE_finite_freq})], yet this effect is subdominant compared to the conservative tidal torque near mode resonance [Eq.~(\ref{eq:dphi_diss_vs_dphi_cons})]. 

(3). We computed the tidal phase shift as functions of both frequency and time using both energy-balancing arguments and osculating orbits (Sec.~\ref{sec:time_phase_shift}). The dominant effect of mode resonance is a conservative energy transfer from orbit to the NS mode [Eq. (\ref{eq:g_phi_int})]. Consistent with the previous analysis, this effect can be approximated by a sudden change in the waveform frequency [Eqs.~(\ref{eq:Delta_phi_vs_f_apprx}) and (\ref{eq:Delta_phi_vs_t_apprx})]. The effect scales linearly with the amount of phase shift at mode resonance, which can be 0.5 and 10 radians for the $l=3$ and $l=2$ f-modes, respectively (top panels in Figs. \ref{fig:dphi_l3} and \ref{fig:dphi_l2}). 
While our analytical approximations of the phase shifts have good accuracy for the $l=3$ f-mode (Fig.~\ref{fig:dphi_l3}), for the $l=2$ f-mode linear theory is insufficient, as illustrated in Fig.~\ref{fig:dphi_l2};  we discuss this point below.  
We further considered the dissipative effect due to the interaction of NS and orbital mass quadrupoles and showed the finite frequency correction cannot be captured with the effective Love number [Eq. (\ref{eq:dE_finite_freq})]. The dynamical tide has no net contribution to this effect [Eq. (\ref{eq:dphi_Qns_dyn_Qorb})]. 

(4). Further signatures associated with the dynamical tide were examined in Secs~\ref{sec:ecc} and \ref{sec:GW_f_mode}. In particular, the orbit cannot remain quasi-circular after a mode's resonance. The eccentricity excited by the dynamical tide can further cause the frequency to be non-monotonic with time (Fig.~\ref{fig:ecc_l2}). The excited f-mode can also emit GW on its own. Such a signal can be detected with an SNR greater than unity at 50 Mpc with the next generation of GW detectors for a wide range of NS spin (Fig. \ref{fig:h_ns}). 

While we focused on the resonant excitation of f-modes, our approach is general and can be applied to the resonant excitation of other NS modes \cite{Lai:94c, Reisenegger:94,  Yu:17a, Yu:17b, Kuan:21, Kuan:21b, Ho:99, Xu:17, Poisson:20, Ma:21, Gupta:21, Tsang:12, Pan:20, Passamonti:21}. For example, the equilibrium component of each mode would allow us to compute the correction to the effective Love number from other NS modes (similar to, e.g.,  Ref.~\cite{Andersson:21}). Our Eq.~(\ref{eq:b_a_eq}) enables the incorporation of finite-frequency corrections while avoiding divergence near resonance. On the other hand, our analysis in Sec.~\ref{sec:ecc} proves that the phase shift near mode resonance [Eqs.~(\ref{eq:Delta_phi_vs_f}) and (\ref{eq:Delta_phi_vs_t})] is the dominant dynamical tide effect of low-frequency modes (such as gravity modes) whereas the post-resonance eccentricity excited by such a mode is subdominant [Eq.~(\ref{eq:ecc_vs_dphi})]. 

As a caveat, our analysis adopted a few simplifying assumptions. In particular, our treatment of the orbital dynamics is at the lowest order (i.e., Newtonian). Since the f-mode excitation mainly happens during the late inspiral stage, relativistic corrections are crucial and we plan to upgrade our study to the EOB framework~\cite{Steinhoff:16, Hinderer:16, Steinhoff:21} to generate waveforms to be used for data analysis. 
Of particular significance to the dynamical tide are the corrections to the resonance frequency produced by redshift and frame dragging (see e.g., fig. 1 of Ref.~\cite{Steinhoff:21}). These effects further determine the critical spin rate required for the f-modes to experience resonance. The amplitude of the mode further depends on the duration of resonance through the $\sqrt{1/\dot{\omega}_r}$ factor [e.g., Eq.~(\ref{eq:c_a_dyn})], and in the late inspiral stage the orbital decay rate can see significant corrections from the higher PN terms. 
The path forward to upgrading our analysis to the EOB formulation is well-defined. Note that our tidal Hamiltonian $H_{\rm mode} + H_{\rm int}$ plays the same role as  $H_{\rm DT}$ in Ref.~\cite{Steinhoff:21}, which also accounts for relativistic redshift and frame dragging. The total Hamiltonian, the central part of the EOB formulation, is constructed when we further sum the tidal Hamiltonian to the upgrade PP Hamiltonian, which can be found in, e.g., Ref.~\cite{Khalil:20}. More specifically, our $\kappa_l$ can directly replace the effective Love number in Ref.~\cite{Steinhoff:21}, while the $\gamma_l$ [Eq.~(\ref{eq:gamma_l})] term describes a back-reaction torque missing in previous studies that should be further integrated into the EOB evolution equations. Besides the conservative dynamics, our Eqs.~(\ref{eq:dE_finite_freq}) and (\ref{eq:dE_dyn}) further provide the finite-frequency and dynamical tide corrections to the dissipative GW radiation from the interaction between NS and orbital quadrupoles, which improves the existing EOB models that compute the dissipative effects under the adiabatic limit [see, e.g., the discussion below eq. (5) of Ref.~\cite{Hinderer:16}].

At the same time, higher-order spin corrections of the NS should be incorporated for the rapidly spinning NSs considered here. We note that spin corrects the effective Love number [Eq.~(\ref{eq:k_lm_eq_omega_0})] starting at $\Omega^2$ due to the finite frequency response of the modes. The modification to the background structure also enters at the same order~\cite{Hartle:67}, yet such an effect is ignored in the current study. Further investigations are hence required.

Our Fig.~\ref{fig:dphi_l2} highlights the need for developing tidal theories in spinning NSs beyond the linear order in both $(\Delta r/r)$ and $(\xi/R)$. The higher-order effect in $(\Delta r/r)$, which corresponds to higher-order back reactions formally entering at $(R/r)^{10}$, can be captured if we do not use a closed-form expression for the tidal phase shift but instead numerically solve the coupled differential equations governing the evolution of tide and orbit. For computational efficiency, a hybrid approach may be taken where analytical expressions can be when $u\lesssim -1$ [Eq.~(\ref{eq:u_def})], and only the evolution with $u\gtrsim-1$ is tracked numerically. 

To capture higher-order hydrodynamical effects [i.e., higher-order effects in ($\xi/R$)], new physical ingredients must be included. At the lowest order, this includes adding a three-wave interaction in the Hamiltonian [Eq.~(\ref{eq:Hamiltonian})] and a nonlinear correction to the mass quadrupole [Eq.~(\ref{eq:multipole_from_amp})]; see Ref.~\cite{Yu:23a} for details. The nonlinear hydrodynamic effects enter at $(R/r)^{8}$ and therefore can be more significant than the higher-order back-reaction effects. Indeed, as shown in Ref.~\cite{Yu:23a}, the nonlinear hydrodynamic effect effectively lowers the eigenfrequency of a mode, enabling f-mode resonance to happen in (more realistic) NSs with milder spins. 
Moreover, the anharmonicity of the f-mode~\cite{Landau:82}, while being a higher order effect in $(R/r)$, can become import as it is amplified by mode resonance. 
The $l_a=2$ f-mode can reach an energy nearly 10\% of the binding energy with resonance, which suggests the effective natural frequency of the f-mode would be lower than the linear value by $\sim 10\%$~\cite{Yu:23a}. Such a correction can become comparable or even exceed various relativistic corrections considered in Ref.~\cite{Steinhoff:21}. 
The peak frequency of the GW signal from a resonantly excited f-mode (Fig.~\ref{fig:h_ns}) will also shift due to the nonlinear hydrodynamic corrections (cf. the anharmonicity of an oscillator~\cite{Landau:82}). 
Another type of nonlinear hydrodynamic effect is the excitation of small-scale fluid instabilities such as the p-g instability \cite{Weinberg:13, Venumadhav:14, Weinberg:16, Essick:16, GW170817pg}. As the tide deviates more from the adiabatic limit and reaches higher energy in a spinning NS, cancellations in the instabilities growth rate may be reduced, potentially amplifying their impact on the GW signal.  
Future investigations along these directions are therefore crucial. 

The strong $l=2$ tide in the presence of f-mode resonance also means we need to be cautious when building frequency-domain phenomenological models including tides (e.g., Ref.~\cite{Dietrich:19}). As shown in Fig.~\ref{fig:ecc_l2}, the quasi-circular approximation is no longer accurate and the frequency evolution is no longer monotonic after the $l=2$ f-mode's excitation. In the example shown in Fig.~\ref{fig:ecc_l2}, the orbit hits $f=1450\,{\rm Hz}$ at 4 different instants of time. As a result, the orbit is not uniquely defined at a given frequency, violating the underlying assumption of many frequency-domain phenomenological models. Nonetheless, approaches similar to that proposed in, e.g., Ref.~\cite{Yunes:09} may be applied to handle eccentricity in the frequency domain. 

We do not consider the impact of NS crust. Ref.~\cite{Passamonti:21} showed that the presence of a crust does not affect the f-mode we consider here. We nonetheless note that the energy stored in the f-mode is large enough to potentially break the crust \cite{Tsang:12, Pan:20}. If this does happen, its impact on the GW signal (both the tidal phase shift and the GW emission from the f-mode itself) is unclear and requires further investigation. 

Besides the tidal phase shift considered in this work, a rapidly spinning NS produces other matter effects in the inspiral stage including dephasing and precession (in generic spin configurations) due to spin-induced quadrupole, see, e.g., Ref~\cite{Lyu:23} and references therein. The post-merger oscillation is yet another signal due to matter effects with promising detection prospects. A complete waveform model should coherently integrate all these components to maximize the information that can be extracted. 

Suppose a misaligned spin is produced because of the dynamical formation of the binary. In that case, there is also a possibility that the binary will have some residual eccentricity when it enters the sensitivity band of a ground-based GW detector \cite{Seto:13, East:13, Rodriguez:15, Naoz:16, Samsing:17, Antonini:17}.  Discussions of tides in eccentric BNS systems with varying initial eccentricities can be found in Refs.~\cite{Gold:12, Chirenti:17, Yang:18, Vick:19b, Parisi:18, Yang:19, Wang:20}. 
Interestingly, Ref. \cite{Gamba:23} showed that f-mode excitations observed in numerical relativity simulations are not captured with the effective Love number prescription. This is consistent with our finding because the solutions constructed following Refs. \cite{Steinhoff:16, Hinderer:16, Steinhoff:21} do not correctly capture the oscillation frequency of the NS mass multipoles in the post-resonance regime (Fig.~\ref{fig:Q33}). A decomposition of the mass multipoles into an equilibrium (following the orbital phase) and a dynamical component (corresponding to the excited oscillator) similar to our Eq.~(\ref{eq:mode_decomposition}) (see also Ref.~\cite{Arras:23}) is instead a promising way of modeling tides in eccentric systems. Such a decomposition can also help separate terms that depend only on the instantaneous orbital configuration and those depending on evolution history.

\begin{acknowledgments}
We thank Neil Cornish and Rosemary Mardling for useful discussions during the conceptualization and preparation of this work. 
H.Y. acknowledges support from NSF grant No. PHY-2308415. P.A. and N.W. acknowledge support from  NSF grant No. AST-2054353.
\end{acknowledgments}

\appendix


\bibliography{ref}

\end{document}